%% file: main.tex
\documentclass[11pt,a4paper]{article}

\usepackage[
  top=0.9in, bottom=0.9in, left=0.85in, right=0.85in,
  headheight=28pt, headsep=14pt
]{geometry}

\usepackage[utf8]{inputenc}
\usepackage[T1]{fontenc}
\usepackage{libertine}
\usepackage[varqu,scaled=0.88]{zi4}

\usepackage{deepinternet}

\usepackage{setspace}
\usepackage{fancyhdr}
\usepackage{float}
\usepackage{placeins}

\graphicspath{{figures/}}

\title{%
  \vspace{-1cm}%
  \textbf{\Huge \sys}\\[10pt]
  \textbf{\LARGE The Secure and Verifiable Interoperability Protocol \\for An Internet of Agents}\\[12pt]
  \large\textit{A Paper from the DeepKernel Lab}%
}

\author{%
  Zhenhua Zou\quad
  Sheng Guo\quad
  Qiuyang Zhan\\[3pt]
  Lepeng Zhao\quad
  Shuo Li\quad
  Zhuotao Liu\textsuperscript{\dag}\\[6pt]
  {
  \textsuperscript{\dag}Corresponding author:
  \texttt{zhuotaoliu@tsinghua.edu.cn}}
}

\date{}

\begin{document}

\maketitle
\thispagestyle{fancy}

\input{sections/abstract}

\input{sections/introduction}

\input{sections/background}

\input{sections/overview}

\input{sections/identity}

\input{sections/registration}

\input{sections/authentication}

\input{sections/accountability}

\input{sections/analysis}

\input{sections/related}

\input{sections/discussion}

\input{sections/conclusion}

\bibliographystyle{unsrt}
\bibliography{bibliography/refs}

\end{document}

%% file: sections/abstract.tex
\begin{abstract}
The emerging Internet of Agents---a global environment in which
LLM-powered agents discover peers, negotiate trust, invoke tools, and
delegate tasks across organizational boundaries---is currently being
standardized from the communication layer upward. Protocols increasingly
specify how agents exchange messages, but not how an agent proves what it
is, what it is authorized to do, whether its advertised capabilities are
genuine, or how its actions remain accountable after delegation. We argue
that this leaves the ecosystem without the foundation for \emph{secure
interoperability}.

We present \sys, a trust-native protocol suite that supplies this trust
substrate alongside existing Internet and agent communication protocols. 
\sys is organized into four layers---Persistent
Identity~(L0), Discovery~(L1), Trust Negotiation~(L2), and
Accountability~(L3)---covering nine security aspects across the agent
lifecycle. Its core contribution is a set of four layer-aligned design
primitives. First, Agent Identity Cards provide persistent agent identity
with four-dimensional binding across developer, code package, operator, and
deployment context. Second, capability-aware discovery turns skill and tool
advertisements into DID-bound Verifiable Credential manifests, so discovery
results are verified for issuer provenance, subject binding, permission
alignment, and freshness before interaction begins. Third, trust negotiation
combines monotonic capability attenuation with two-tier access control,
making least privilege a signed structural invariant while preserving
application-level policy independence. Fourth, kernel-mediated
cryptographic audit trails bind usage, delegation, and execution traces to
agent identity without requiring a consensus ledger.

\sys does not replace existing agent protocols such as MCP, A2A, ANP, or AG-UI; it defines the missing
trust-relevant primitives that those protocols need: AIC capability boundaries,
DID-bound manifest VCs, least-privilege session tokens, and signed execution traces. 
This separation ensures that communication protocols can evolve
independently, while trust semantics remain explicit, portable, and
verifiable. We compare \sys against 50+ related efforts across agent
protocols, decentralized identity, OAuth/OIDC extensions, zero-trust
governance, delegation systems, and audit architectures. 
We show that \sys is the only 
architecture that jointly enforces persistent identity,
capability-aware discovery, trust negotiation, and accountability as a single four-layer trust substrate for the Internet of Agents.
\end{abstract}

%% file: sections/introduction.tex
\section{Introduction}
\label{sec:intro}

The emergence of LLM-powered autonomous agents marks a qualitative shift in
how software interacts with the world. Unlike traditional microservices that
execute deterministic APIs, these agents reason over natural language, invoke
tools dynamically, and collaborate with other agents to accomplish complex,
multi-step tasks~\cite{react,autogen,metagpt,ioa-weaving}. Industry analysts
project that by 2029 agentic AI will autonomously resolve a large share of
common customer service issues without human intervention~\cite{gartner-agentic}, and major cloud providers
have already released agent-to-agent communication
protocols~\cite{a2a-protocol,mcp-spec,anp-whitepaper} to enable such
collaboration at scale.

This trajectory evolves toward an \emph{Internet of Agents}
(IoA)~\cite{ioa-weaving,ioa-fundamentals,agent-osi}---a global network where
billions of heterogeneous agents discover one another, negotiate task
parameters, exchange value, and compose into ad-hoc workflows across
organizational boundaries. Multiple research groups have begun designing
protocol stacks for this vision: Fleming et al.\ propose reference layers for
agent communication (L8) and semantics (L9) atop TCP/IP~\cite{ioa-layered}; the Agent-OSI
project outlines a six-layer reference architecture including settlement and
provenance~\cite{agent-osi}; and comprehensive surveys catalog the growing
zoo of protocols spanning agent-to-agent (A2A), agent-to-tool (MCP), and
related interaction axes~\cite{agent-protocol-survey,agent-protocol-survey-ehtesham,acps}.\footnote{In Ehtesham et al., ``ACP'' denotes the Agent Communication Protocol (REST/HTTP), not the Agent Client Protocol~\cite{acp-protocol}.}

\parab{The missing trust layer}
A striking pattern emerges across these efforts: they focus predominantly on
\emph{how agents communicate} where security is again treated as the second-class citizen. Yet the agent threat landscape is qualitatively different from the traditional
networking and computer systems. Agents operate with delegated authority, make
non-deterministic decisions, and dynamically compose into workflows whose
authorization requirements cannot be statically pre-computed. An agent that
impersonates a supply-chain optimizer, escalates its capabilities beyond
delegation scope, or repudiates a financial commitment can cause cascading
damage that no transport-layer encryption or OAuth token can
prevent~\cite{agentrfc,anbiaee-threat-model,safe-ioa}. A first wave of
\emph{trust-layer} proposals has begun to address this gap---Microsoft's
AgentMesh~\cite{agentmesh}, which its documentation describes as ``SSL for AI agents'';
Huang et al.'s unified zero-trust architecture for the agentic
web~\cite{zt-unified}; AIP~\cite{aip-protocol}, which fuses identity,
attenuation, and provenance into invocation-bound capability tokens; and
Ramachandran and Mishra's enterprise identity-aware governance
framework~\cite{identity-aware-governance}. These efforts converge on the
diagnosis but diverge in mechanism: each enforces least-privilege through
\emph{runtime} policy evaluation, behavioral attestation, or per-call
Datalog rather than through structural invariants signed into the agent's
credential at issuance. In the language of classical protocol design, the
IoA still lacks its \emph{trust plane}---the security substrate that binds
identity, authorization, and accountability into a coherent whole that
holds even when downstream policy engines, attestors, or operators are
misconfigured or compromised.

\parab{Lessons from our prior work}
In \blocka~\cite{blocka2a}, we presented a unified trust framework
for multi-agent systems, integrating decentralized identifiers (DIDs),
blockchain-anchored ledgers, and smart-contract-enforced access control. The
framework demonstrated that security \emph{can} be systematically layered
onto agent collaboration protocols such as A2A~\cite{a2a-protocol}. However, its
reliance on blockchain infrastructure introduced deployment constraints:
on-chain transaction latency, gas costs, and the requirement for a shared
ledger among all participants limited applicability to environments where
blockchain nodes are available and economically viable. The core security
\emph{ideas}---cryptographic agent identity, capability-bounded access
control, immutable audit trails, and defense orchestration---remain sound,
but their \emph{realization} must be generalized beyond any specific type of 
infrastructure.

\parab{This paper: \sys}
We present \sys, the first trust-native protocol suite for secure and verifiable agent-to-agent interoperability. 
Rather than proposing another communication stack, \sys
provides the \emph{trust layers} that any communication protocol
(MCP, A2A, or future designs) can build upon. Its four layers address
nine critical aspects of agent-native security:

\begin{enumerate}[leftmargin=2em,itemsep=2pt]
  \item \textbf{Layer~0 --- Agent Identity:} \emph{persistent identity} via cryptographic agent
    identity cards (AICs) with four-dimensional binding to developer, code
    package, operator, and deployment context.
  \item \textbf{Layer~1 --- Discovery:} \emph{capability-aware discovery} in which
    each skill and tool is a signed Verifiable Credential (VC) bound to the
    agent's DID. Manifest VCs are presented during discovery and verified for
    supply-chain authenticity, subject binding, and permission alignment.
  \item \textbf{Layer~2 --- Trust Negotiation:} \emph{authentication} via mutual attestation,
    \emph{delegation} via chains with monotonic capability attenuation, and a two-tier
    \emph{access control} model.
  \item \textbf{Layer~3 --- Accountability:} \emph{token-usage tracing}, \emph{payment}
    primitives, and \emph{action accountability} via identity-signed execution traces
    and non-repudiation.
\end{enumerate}

\sys is guided by five design principles (detailed in
\S\ref{subsec:principles}) and introduces four novel design primitives.
We summarize each below using a uniform pattern---shared goal with prior
work, the specific divergence, and the operationally observable
consequence---a pattern we apply throughout \S\ref{sec:related} when
contrasting \sys with related systems.

\begin{itemize}[leftmargin=2em,itemsep=2pt]
  \item \textbf{Persistent Identity (L0).} An Agent Identity Card (AIC)
    cryptographically binds four identity dimensions---developer, code
    package, operator, and operational context---into a single verifiable
    credential. The goal of giving each agent a verifiable persistent
    identity is shared with SPIFFE workload identities, OAuth~2.0 client
    credentials, W3C DIDs, OIDC for Agents (OIDC-A)~\cite{oidc-a}, the
    OpenID Foundation's strategic agenda for agentic
    IAM~\cite{iam-agentic-ai}, and AIP's capability
    tokens~\cite{aip-protocol}; the divergence is that each of those
    models binds only a single dimension---workload instance, client
    application, holder key, or token issuer---whereas the AIC binds all
    four dimensions independently and signs their conjunction. The
    consequence is that an attacker who compromises the operator's
    deployment cannot impersonate a different developer or substitute a
    different code package, because each dimension carries its own
    independently verifiable signature.
  \item \textbf{Capability-Aware Discovery (L1).} Each skill and tool an
    agent advertises is a signed Verifiable Credential (VC) whose
    \texttt{credentialSubject.id} is the agent's own DID---making it
    non-transferable and replay-resistant. Skill VCs are issued by skill
    distributors, tool VCs are issued by tool providers, and both can carry
    optional GAR endorsements before being presented in the agent's
    registration record during discovery. The goal of
    making agents discoverable is shared with ANP's DID-based
    discovery~\cite{anp-whitepaper}, A2A's well-known Agent
    Cards~\cite{a2a-protocol}, and registry-style agent catalogs; the
    divergence is that \sys treats discovery results as \emph{verified
    capability credentials} rather than self-declared descriptions.
    Requesters verify each manifest VC for supply-chain authenticity,
    subject binding to the presenter's AIC, and permission alignment
    against the agent's capability boundary~$\Smax$. The consequence is
    that discovery cannot become an implicit escalation channel: an
    agent cannot advertise skills it does not hold, tools it has not been
    authorized to use, or permissions it was not granted, because every
    claim is cryptographically bound to the agent's identity and
    independently verifiable at discovery time.
  \item \textbf{Trust Negotiation (L2).} \sys combines a four-stage
    monotonic attenuation chain (Developer $\to$ Operator $\to$ GAR
    $\to$ Runtime) with a two-tier access control model that separates
    infrastructure-tier cryptographic verification from application-tier
    declarative policy. The goal of bounded authorization is shared with
    AIP's Biscuit/Datalog attenuation~\cite{aip-protocol}, Saavedra's
    Delegation Grants~\cite{saavedra-delegation}, AgentMesh's policy
    plane~\cite{agentmesh}, OAuth~2.0 token exchange~\cite{rfc8693}, and
    enterprise governance frameworks~\cite{identity-aware-governance};
    the divergence is that \sys encodes the capability boundary into
    signed credentials and then evaluates application policy on an
    independent path. The consequence is that no policy edit, JIT
    decision, or misconfigured PDP downstream of issuance can grant
    capabilities the preceding stage did not authorize, while an
    application-tier regression still cannot weaken cryptographic
    identity verification.
  \item \textbf{Accountability (L3).} Every agent
    action is recorded in a tamper-evident hash chain and signed by the
    agent's kernel-protected private key. The goal of immutable
    accountability is shared with blockchain-anchored ledgers
    (like \blocka~\cite{blocka2a}) and centralized audit logs; the
    divergence is that \sys achieves non-repudiation via trusted
    kernel-mediated cryptography rather than distributed consensus or
    trusted third parties. The consequence is that agents can prove
    their execution history and attribute costs across boundaries in
    completely decentralized or resource-constrained environments where
    global ledgers are unviable.
\end{itemize}

\parab{Scope and positioning}
\sys is a \emph{positioning paper}: it presents the conceptual framework,
protocol architecture, and design rationale at a level suitable for guiding
subsequent protocol specifications, formal verification, and systems
implementation. Detailed cryptographic proofs, performance benchmarks, and
production deployment experiences are deferred to companion publications
currently in preparation.

\parab{Contributions}
In summary, this paper makes the following contributions:
\begin{enumerate}[leftmargin=2em,itemsep=2pt]
  \item We articulate the case for \emph{secure interoperability} as the
    foundational missing layer in the Internet of Agents, distinct from
    communication interoperability and grounded in a dedicated trust
    substrate.
  \item We present \sys, a four-layer security protocol suite that
    coherently addresses nine security aspects: persistent identity,
    capability-aware discovery, authentication, delegation, access
    control, payment, semantic tagging, token-usage tracing, and action
    accountability. The first five receive detailed protocol-level
    treatment; the latter four are specified as framework-level
    primitives that establish architectural slots for future refinement.
  \item We introduce four novel design primitives: the Agent Identity
    Card for persistent identity, capability-aware discovery through
    DID-bound Verifiable Credential manifests that make skill and tool
    claims cryptographically verifiable at discovery time, trust
    negotiation combining the monotonic capability attenuation chain
    with the two-tier A2A access control model, and kernel-mediated
    cryptographic audit trails for accountability.
  \item We position \sys within the current landscape of agent protocol,
    identity, and governance efforts (\S\ref{sec:related})---organized
    into six clusters spanning IoA stacks and surveys, agent IAM and
    OAuth/OIDC extensions, zero-trust DID/VC architectures,
    delegation-centric protocols, security-design principles, and
    industry trust overlays---and make the joint-coverage argument
    explicit: each individual primitive in \sys is anticipated by some
    prior work, but the conjunction of persistent identity,
    capability-aware discovery, trust negotiation with monotonic
    attenuation and two-tier access control, and kernel-mediated
    accountability appears in
    no prior single architecture (\Cref{tab:comparison}).
\end{enumerate}

The remainder of this paper is organized as follows.
\S\ref{sec:background} surveys the agent interaction landscape and threat
model. \S\ref{sec:overview} presents the design philosophy and protocol suite
overview. \S\ref{sec:identity}--\S\ref{sec:accountability} detail each
layer. \S\ref{sec:analysis} provides a security analysis and comparison with
existing approaches. \S\ref{sec:related} discusses related work in depth.
\S\ref{sec:discussion} outlines limitations and future directions.
\S\ref{sec:conclusion} concludes.

%% file: sections/background.tex
\section{Background \& Threat Landscape}
\label{sec:background}

\subsection{The Agent Interaction Landscape}
\label{subsec:agent-landscape}

Modern agent-to-agent communication is shaped by a rapidly growing set of
protocols, each targeting a different facet of the problem:

\parab{Model Context Protocol (MCP)}
The Model Context Protocol (MCP)~\cite{mcp-spec}, introduced by Anthropic and
now maintained as an open, vendor-neutral specification, standardizes how an
agent (client) discovers and invokes external tools hosted by MCP servers via
JSON-RPC. Third-party deployment surveys report over 97 million monthly SDK
downloads as of early 2026~\cite{mcp-production}; MCP is widely adopted for
agent-to-tool integration. The core specification, however, does not define
agent-native identity, mutual authentication between agents, or authorization
semantics beyond transport-level security.

\parab{Agent-to-Agent Protocol (A2A)}
The Agent-to-Agent (A2A) protocol~\cite{a2a-protocol}, developed under the
A2A open specification effort, enables peer-to-peer task delegation between
agents using capability-based \emph{Agent Cards}. A2A supports both
synchronous and asynchronous interactions and provides structured task
lifecycle management. Published security guidance centers on HTTPS and
OAuth~2.0; the specification does not yet standardize agent-native identity,
structured delegation chains, or cross-domain trust comparable to a dedicated
trust plane.

\parab{Agent Network Protocol (ANP)}
ANP~\cite{anp-whitepaper} introduces a three-layer architecture---identity
and encrypted communication, meta-protocol negotiation, and application
protocol---using W3C DIDs and JSON-LD for open agent discovery. ANP focuses
on the ``agentic web'' vision but, in its published white paper, treats
access control and end-to-end accountability less extensively than
identity and discovery.

\parab{Emerging protocol stacks}
Several groups propose higher-level architectures. Fleming
et al.~\cite{ioa-layered} propose reference layers for agent communication
(L8) and agent semantics (L9) above TCP/IP (not part of the classical OSI
stack). Agent-OSI~\cite{agent-osi} proposes a six-layer
decentralized stack including settlement and provenance. ACPS~\cite{acps}
defines registration, discovery, interaction, and tooling protocols.
Coral Protocol~\cite{coral-protocol} provides open infrastructure for
agent communication, coordination, trust, and payments framed as the
``Internet of Agents.'' Surveys~\cite{agent-protocol-survey,agent-protocol-survey-ehtesham,
agentic-ai-frameworks} catalog and compare these efforts.

\parab{Beyond A2A: adjacent interaction axes and trust overlays}
The protocol space continues to multiply along distinct interaction axes
that complement A2A: AG-UI~\cite{ag-ui-protocol} for agent-to-user
streaming, the Agent Client Protocol (ACP)~\cite{acp-protocol} for
editor-to-coding-agent flows, IBM ContextForge~\cite{ibm-contextforge}
as a federated gateway proxying MCP/A2A/REST, and the
\texttt{agents.json} specification~\cite{agents-json} extending OpenAPI
with agent-specific interaction contracts. Most recently, Microsoft's
AgentMesh~\cite{agentmesh} describes itself in project documentation as
``SSL for AI agents,''
layering SPIFFE/SVID workload identity, a runtime policy engine, and
A2A/MCP/IATP protocol translators atop existing communication stacks; a
parallel academic effort, Huang et al.'s unified zero-trust
architecture~\cite{zt-unified}, adds behavioral attestation and trust-adaptive runtime environments to the same trust-overlay paradigm. We
return to both as our primary industry and academic comparators in
\S\ref{sec:related}.

\begin{insight}[Observation]
Existing protocols and stacks overwhelmingly prioritize \emph{communication
interoperability}: how agents find each other, exchange messages, and
invoke tools. Security is treated as an aspect to be satisfied at each
layer rather than as a \emph{foundational design center} with its own
coherent architecture. Even the most security-conscious trust overlays
(AgentMesh~\cite{agentmesh}, the unified zero-trust
architecture of Huang et al.~\cite{zt-unified}) enforce least-privilege
through runtime policy evaluation and behavioral attestation rather than
through monotonic capability attenuation that holds at the credential
level---so a misconfigured policy or a compromised attestor can silently
widen capability. This leaves critical gaps that individual protocol
patches cannot close.
\end{insight}

\subsection{Agent-Specific Threat Model}
\label{subsec:threat-model}

The agent ecosystem introduces threats that are qualitatively different from
traditional web services. We identify six categories that motivate the design
of \sys.

\begin{table}[t]
\centering
\caption{Agent-specific threat categories and their structural causes.}
\label{tab:threats}
\small
\begin{tabularx}{\textwidth}{@{}l >{\raggedright\arraybackslash}X >{\raggedright\arraybackslash}X@{}}
\toprule
\textbf{Threat} & \textbf{Description} & \textbf{Structural Cause} \\
\midrule
T1: Identity spoofing &
  An agent claims to be built by a trusted developer or to possess capabilities it lacks. &
  Agents use self-declared names or API keys; no binding to developer, code, or operator. \\[4pt]
T2: Capability escalation &
  A child agent exercises permissions beyond what its parent delegated. &
  Static permission models lack monotonic attenuation; delegation is often unconstrained. \\[4pt]
T3: Delegation abuse &
  Unbounded delegation chains create laundering paths for authority. &
  No depth bounding, no cascading revocation, no tenant isolation in delegation. \\[4pt]
T4: Cross-domain trust breakdown &
  Agents from different organizations cannot verify each other's identity or capabilities. &
  No shared trust anchor; each framework uses its own identity model. \\[4pt]
T5: Payment \& usage fraud &
  An agent consumes LLM tokens or services without traceable identity. &
  Token usage is tied to API keys, not to cryptographic agent identity; no metering protocol. \\[4pt]
T6: Action repudiation &
  An agent denies having performed a high-impact action (e.g., financial transfer). &
  Execution traces are not cryptographically signed; audit logs are mutable. \\
\bottomrule
\end{tabularx}
\end{table}

\Cref{tab:threats} summarizes these threats. Prior threat
analyses~\cite{agentrfc,anbiaee-threat-model,safe-ioa,blocka2a} have
identified overlapping subsets; our contribution is to map them to their
\emph{structural causes}---the architectural deficiencies that enable
each threat---which directly motivates the layered design of \sys.

The taxonomy in \Cref{tab:threats} is not constructed in isolation. It
synthesizes three independent lines of analysis: Anbiaee
et al.~\cite{anbiaee-threat-model} compare MCP, A2A, Agora, and ANP and
identify twelve protocol-level risks, several of which (cross-protocol
credential laundering, executable-component attestation, lifecycle
risks) map directly onto T1--T4. AgentRFC~\cite{agentrfc} formalizes
eleven security principles as TLA+ invariants and introduces the
Composition Safety principle---the observation that properties holding
for individual protocols can break under composition---which underlies
T2 and T4. Identity-aware governance
analyses~\cite{identity-aware-governance} ground these abstract risks
in production incidents reported in that survey (e.g., the ClawHavoc
supply-chain attack on 824+ malicious agent skills, Cisco's finding that
26\% of analyzed agent skills contain security vulnerabilities, and
30,000+ internet-exposed OpenClaw instances), making T1--T6 verifiable
rather than purely analytical.

\subsection{Why Traditional Internet Security Falls Short}
\label{subsec:why-traditional-fails}

The Internet's existing security mechanisms were designed for human users
interacting with web services. Three fundamental mismatches arise when they
are applied to autonomous agents.

\parab{OAuth/OIDC assumes human principals}
OAuth~2.0~\cite{oauth2} and OpenID Connect~\cite{oidc} model a three-party
flow: a human user authorizes a client application to access a resource
server. Agent-to-agent interactions invert this model---both parties are
non-human, may be ephemeral, and require mutual (not unilateral)
authentication. The OpenID Foundation's strategic agenda for agentic
IAM~\cite{iam-agentic-ai} explicitly catalogs these gaps and surveys
candidate extensions; OIDC-A~\cite{oidc-a} is the most concrete
extension to date, defining standard claims for representing agent
identity, attestation, and delegation chains within OAuth. Yet such
extensions still inherit OAuth's interactive-flow assumptions and bind
to a single \emph{client} dimension, not to the developer, code
package, operator, and operational context that agent provenance
requires~\cite{authenticated-delegation,auth-taxonomy}.

\parab{TLS provides transport, not identity semantics}
TLS~\cite{tls13} secures the channel but does not tell the recipient
\emph{who built} the agent, \emph{what code} it runs, or \emph{who
authorized} its deployment. A TLS certificate binds a public key to a
domain name, not to a developer-code-operator triple. Mutual TLS (mTLS)
adds client-side certificates but still does not express agent
capabilities, delegation depth, or revocation cascades. Recent
zero-trust frameworks add SPIFFE/SVID workload identity above
mTLS~\cite{agentic-ai-identity-security,spiffe-agents,zt-agentic-ai}, which
removes static API keys and enables short-lived attestation, but binds
keys to \emph{workload instances}---a single dimension---rather than to
the conjunction of developer, code package, operator, and operational
context that agent trust decisions require.

\parab{X.509 does not bind agent semantics}
X.509 certificates~\cite{x509} bind a public key to a subject name and
are widely used in PKI. However, agent identity requires richer
semantics: capability boundaries, delegation chains, code-package
hashes, and operational context. Encoding these in X.509 extensions is
technically possible but semantically awkward and incompatible with the
existing CA ecosystem. SPIFFE/SPIRE~\cite{spiffe-agents} provides
workload identity for Kubernetes but, as noted above, does not model
agent-native concepts such as delegation depth, capability attenuation,
or developer-code provenance.

\begin{insight}[Gap]
No existing Internet security mechanism provides the combination of
\emph{persistent identity} with four-dimensional binding,
\emph{capability-aware discovery} with verifiable skill and tool
manifests, \emph{trust negotiation} that combines monotonic capability
attenuation with two-tier access control, and \emph{accountability} via
tamper-evident execution traces. \sys fills this gap with a dedicated
security protocol suite.
\end{insight}

%% file: sections/overview.tex
\section{\sys: Design Philosophy \& Protocol Suite Overview}
\label{sec:overview}

\subsection{Design Principles}
\label{subsec:principles}

\sys is guided by five principles distilled from lessons learned in
\blocka~\cite{blocka2a}, the \deepkernel security
kernel, and the broader agent security literature:

\begin{enumerate}[label=\textbf{P\arabic*.},leftmargin=2.5em,itemsep=4pt]

\item \textbf{Trust as a binding layer, not an add-on.}
  Existing IoA stacks~\cite{ioa-layered,agent-osi,acps} design communication
  first and add security externally---via auth headers, gateway proxies, or
  protocol-specific translators~\cite{agentmesh}. \sys inverts this: its
  trust primitives (AIC capability boundaries, manifest VCs, session tokens,
  trace entries) are designed to be \emph{embedded into} existing protocol
  messages. An MCP invocation carries an AIC-bound capability context; an
  ANP discovery response carries manifest VCs; an A2A interaction carries a
  Layer~2 session token; an AG-UI event stream attaches provenance metadata.
  The trust layer \emph{pervades} the communication layer rather than
  wrapping around it.

\item \textbf{Complementary to existing Internet layers.}
  \sys does not replace TCP/IP, HTTP, or application-layer protocols such as
  MCP and A2A. Instead, it provides a \emph{trust overlay} that any
  transport and any agent protocol can bind to. An agent using A2A over HTTPS
  can adopt \sys's identity and access control layers without modifying A2A's
  message format.

\item \textbf{General cryptographic primitives, no infrastructure lock-in.}
  \blocka{} demonstrated the value of blockchain-anchored auditability, but
  at the cost of deployment generality. \sys uses standard Ed25519 signatures,
  a PKI-style Global Agent Registry (GAR), and challenge-response protocols
  that work in any environment---cloud, edge, or air-gapped.
  Specific deployments can optionally layer additional trust anchors
  (TEEs, HSMs, distributed ledgers) without changing the core protocol.

\item \textbf{Deny-by-default, converge-on-strict.}
  Capabilities, permissions, and access are denied unless explicitly granted.
  When multiple policies combine (e.g., developer declaration, operator
  provisioning, runtime policy), the system computes the
  \emph{intersection}, never the union. Authority can only be attenuated,
  never amplified---a property we call \emph{monotonic capability
  attenuation}.

\item \textbf{Structural guarantees over policy-based assurances.}
  Where possible, \sys encodes security properties as structural invariants
  (e.g., a child AIC's capability set is a cryptographically signed subset of
  its parent's) rather than relying on runtime policy engines to enforce
  them. Structural guarantees survive misconfiguration; policy-based
  assurances do not.

\end{enumerate}

\subsection{Protocol Suite Architecture}
\label{subsec:architecture}

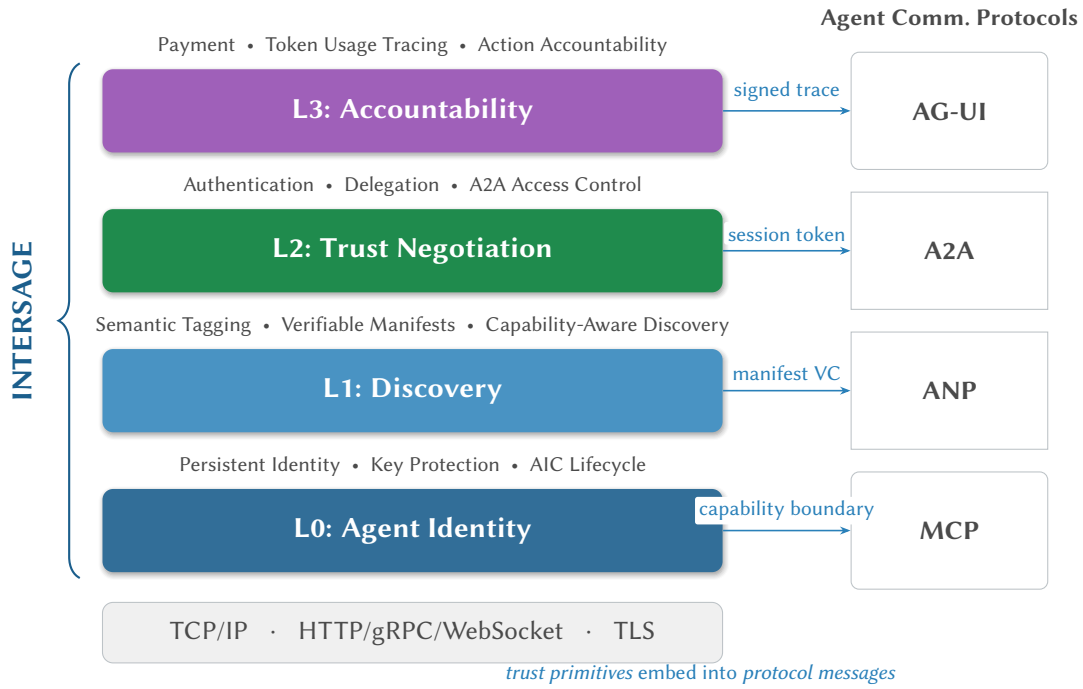
\begin{figure}[t]
\centering
\input{figures/protocol-stack}
\caption{The \sys four-layer protocol suite. Each layer builds on the one
below it and \emph{embeds} trust primitives into existing agent communication
protocols: L0 constrains MCP tool invocation through AIC capability
boundaries, L1 exposes DID-bound manifest VCs through discovery protocols
such as ANP, L2 authorizes A2A interaction through session tokens, and L3
attaches signed execution traces to AG-UI streams.}
\label{fig:protocol-stack}
\end{figure}

\sys is organized into four layers, each addressing a distinct set of
security concerns. \Cref{fig:protocol-stack} illustrates the architecture
and its relationship to the existing Internet protocol stack.

\parab{Layer~0: Agent Identity}
The foundation of \sys. Every agent receives a cryptographic \emph{Agent
Identity Card} (AIC) that binds four identity dimensions---developer, code
package, operator, and operational context---into a single verifiable
credential signed by the Global Agent Registry. Layer~0 also manages the
agent's key lifecycle: provisioning, rotation, and revocation with cascading
guarantees. (\S\ref{sec:identity})

\parab{Layer~1: Discovery}
Before two agents can interact, they must find each other and understand
each other's capabilities. Layer~1 provides capability-aware discovery:
each skill and tool is represented as a signed Verifiable Credential (VC)
bound to the agent's DID, issued by the relevant skill distributor or tool
provider, and presented in the agent's registration record. Requesters
verify each manifest VC for issuer authenticity, subject binding, and
permission alignment against the agent's AIC capability boundary---transforming
directory lookup into a trust-establishing protocol step. Layer~1 supports
both registry-mediated and peer-to-peer discovery modes.
(\S\ref{sec:registration})

\parab{Layer~2: Trust Negotiation}
The interaction layer. When an agent wishes to collaborate with or delegate
to another, Layer~2 orchestrates mutual attestation via challenge-response
over AICs, computes the session capability boundary via intersection, and
evaluates the responder's application-level access policy. Its core
primitive combines monotonic capability attenuation with two-tier access
control: cryptographic credentials define the maximum boundary, while
application policy can only narrow the resulting session. Layer~2 also
manages delegation: a parent agent can issue a child AIC with strictly
attenuated capabilities, forming a cryptographic delegation chain with
bounded depth and cascading revocation. (\S\ref{sec:authentication})

\parab{Layer~3: Accountability}
Every interaction leaves a cryptographic trace. Layer~3 binds LLM
token-usage records to agent identity, provides payment primitives for
agent-to-agent service exchange, and ensures that execution traces are
signed by the agent's kernel-held private key for non-repudiation.
(\S\ref{sec:accountability})

\subsection{Relationship to Existing Protocols}
\label{subsec:relationship}

The agent ecosystem already has a rich set of communication
protocols---MCP, A2A, ANP, AG-UI~\cite{ag-ui-protocol},
ACP~\cite{acp-protocol}---that define \emph{what agents say to each
other}: message formats, task lifecycles, tool invocations, UI event
streams. \sys does not compete with any of them. Instead, it treats
them as \emph{host protocols}---the transport and application channels
into which its trust primitives are embedded---and answers a different
question: \emph{why should agents trust each other?}

We discuss this compositional relationship here, rather than deferring
it to Related Work, because it is integral to the design: every layer
of \sys (\Cref{fig:protocol-stack}) is defined in terms of how its
primitives bind to host-protocol messages, so understanding that binding
is a prerequisite for understanding the protocol suite itself.
The Related Work section (\S\ref{sec:related}) serves a different
purpose: it compares \sys against \emph{security-oriented} architectures
and trust frameworks---AgentMesh, AIP, HDP, ZT-IAM, AgentRFC, among
others---that share \sys's goal of adding trust to the agent ecosystem
but differ in mechanism. Those works are \sys's \emph{comparators};
the protocols below are its \emph{substrates}.

\Cref{fig:protocol-stack} shows this substrate relationship concretely.
\sys operates alongside---not above or below---the host protocols, and
the two concerns compose naturally:

\begin{itemize}[leftmargin=2em,itemsep=2pt]
  \item An MCP tool invocation carries the agent's AIC capability boundary
    (from Layer~0), ensuring that the tool server can verify the caller is
    authorized to invoke the requested tool before execution.
  \item ANP's DID-based discovery responses embed the agent's manifest VCs
    (from Layer~1), enabling any peer resolving an agent's DID to verify
    skill and tool claims via subject binding and permission alignment.
  \item An A2A interaction carries a session token issued by \sys's Layer~2
    after mutual attestation and capability intersection, scoping the
    session to the least-privilege boundary of both parties.
  \item AG-UI events streamed to a frontend include identity-signed
    execution traces (from Layer~3), allowing the UI to display verified
    provenance and non-repudiable action history for each agent.
\end{itemize}

This composability is a direct consequence of Principle~P2: by avoiding
assumptions about the underlying transport or application protocol, \sys
remains agnostic to the rapidly evolving agent communication landscape.

%% file: figures/protocol-stack.tex
\begin{tikzpicture}[
  layer/.style={
    draw=none, rounded corners=4pt, minimum width=8.2cm,
    minimum height=1.1cm, font=\sffamily\bfseries, text=white,
    align=center,
    blur shadow={shadow blur steps=5, shadow xshift=0.5pt,
    shadow yshift=-1pt, shadow blur radius=2pt}
  },
  aspect/.style={
    font=\sffamily\scriptsize, text=didarkgray
  },
  protocell/.style={
    draw=diborder, rounded corners=0pt, minimum width=2.6cm,
    minimum height=1.55cm, font=\sffamily\small\bfseries, fill=white,
    text=didarkgray, align=center, inner sep=3pt
  },
  transport/.style={
    draw=diborder, rounded corners=4pt, minimum width=8.2cm,
    minimum height=0.8cm, fill=gray!12, font=\sffamily\small,
    text=didarkgray
  },
  embed/.style={-{Stealth[length=4pt]}, semithick, color=diaccent},
  embedlbl/.style={font=\sffamily\scriptsize, text=diaccent,
    fill=white, inner sep=1.5pt, rounded corners=1pt},
]

\def\layerht{1.85}
\def\stackx{-0.6}    
\def\protox{6.5}      
\def\gapcol{1.6}      

\node[transport] (transport) at (\stackx, 0)
  {TCP/IP \;\;$\cdot$\;\; HTTP/gRPC/WebSocket \;\;$\cdot$\;\; TLS};

\node[layer, fill=diblue!90]       (l0) at (\stackx, 1.35)              {L0: Agent Identity};
\node[layer, fill=diaccent!85]     (l1) at (\stackx, 1.35+\layerht)     {L1: Discovery};
\node[layer, fill=digreen!80!black](l2) at (\stackx, 1.35+2*\layerht)   {L2: Trust Negotiation};
\node[layer, fill=dipurple!85]     (l3) at (\stackx, 1.35+3*\layerht)   {L3: Accountability};

\node[aspect, above=1pt of l0.north]
  {Persistent Identity \;\textbullet\; Key Protection \;\textbullet\;
   AIC Lifecycle};
\node[aspect, above=1pt of l1.north]
  {Semantic Tagging \;\textbullet\; Verifiable Manifests \;\textbullet\;
   Capability-Aware Discovery};
\node[aspect, above=1pt of l2.north]
  {Authentication \;\textbullet\; Delegation \;\textbullet\;
   A2A Access Control};
\node[aspect, above=1pt of l3.north]
  {Payment \;\textbullet\; Token Usage Tracing \;\textbullet\;
   Action Accountability};

\node[protocell, rounded corners=3pt] (p0) at (\protox, 1.35)            {MCP};
\node[protocell]                      (p1) at (\protox, 1.35+\layerht)   {ANP};
\node[protocell]                      (p2) at (\protox, 1.35+2*\layerht) {A2A};
\node[protocell, rounded corners=3pt] (p3) at (\protox, 1.35+3*\layerht) {AG-UI};


\node[font=\sffamily\footnotesize\bfseries, text=didarkgray,
  above=4pt of p3.north] {Agent Comm.\ Protocols};


\draw[embed] (l0.east) -- (p0.west)
  node[embedlbl, midway, above=2pt] {capability boundary};

\draw[embed] (l1.east) -- (p1.west)
  node[embedlbl, midway, above=2pt] {manifest VC};

\draw[embed] (l2.east) -- (p2.west)
  node[embedlbl, midway, above=2pt] {session token};

\draw[embed] (l3.east) -- (p3.west)
  node[embedlbl, midway, above=2pt] {signed trace};

\draw[decorate, decoration={brace, amplitude=8pt},
  thick, diblue]
  ([xshift=-4.4cm, yshift=-2pt]l0.south)
  -- ([xshift=-4.4cm, yshift=2pt]l3.north)
  node[midway, left=14pt, font=\sffamily\bfseries, text=diblue,
    rotate=90, anchor=south] {\textsc{INTERSAGE}};

\node[font=\sffamily\scriptsize\itshape, text=diaccent, align=center]
  at (3.2, -0.55) {trust primitives \emph{embed into} protocol messages};

\end{tikzpicture}

%% file: sections/identity.tex
\section{Layer~0: Persistent Agent Identity}
\label{sec:identity}

\begin{insight}[Design Thesis]
Without verified identity, every other security property is built on air.
Layer~0 ensures that every agent in the \sys ecosystem is a
\emph{cryptographic principal}---unforgeable, verifiable, and bound to its
provenance.
\end{insight}

\subsection{Agent Identity Card (AIC)}
\label{subsec:aic}

The \emph{Agent Identity Card} is the foundational credential in \sys. An
AIC is a verifiable credential that binds an agent's identity to its
permission boundary. Formally:
\begin{equation}
\label{eq:aic}
\AIC = \Sign_{\GAR}\!\bigl(\DIDag \;\|\; \Kpub \;\|\; \Smax\bigr)
\end{equation}
where $\DIDag$ is a structured agent identifier derived from the
$\langle\textit{developer},\allowbreak\; \textit{code\_pkg},\allowbreak\;
\textit{deploy\_ctx}\rangle$ triple,
$\Kpub$ is the agent's Ed25519 public key (whose corresponding private key
$\Kpriv$ never leaves the isolated trust boundary), and $\Smax$ is the
\emph{capability boundary}---the maximum set of permissions this agent may
ever exercise.

The AIC is signed by the Global Agent Registry (GAR) using its root
Ed25519 key, analogous to a Certificate Authority signing an X.509
certificate. However, unlike X.509, the AIC natively encodes agent-specific
semantics: capability sets, delegation depth, identity assurance levels, and
operational context.

\subsection{Four-Dimensional Identity Binding}
\label{subsec:four-dim}

A persistent agent identity must answer four questions simultaneously:
\emph{who built it}, \emph{what code does it run}, \emph{who deployed it},
and \emph{in what context does it operate}. \sys binds all four into the AIC
through what we call \emph{four-dimensional identity binding}.

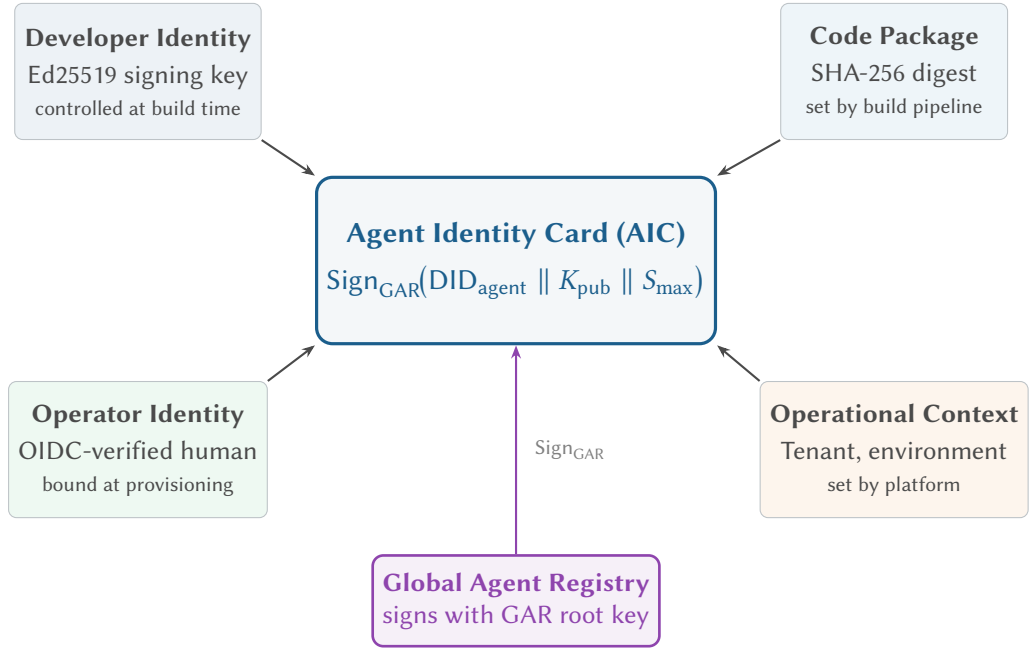
\begin{figure}[t]
\centering
\input{figures/aic-structure}
\caption{Four-dimensional identity binding in the Agent Identity Card. Each
dimension is controlled by a different party and verified independently.}
\label{fig:aic-structure}
\end{figure}

\begin{table}[t]
\centering
\caption{The four identity dimensions of an AIC.}
\label{tab:four-dim}
\small
\begin{tabularx}{\textwidth}{@{}l >{\raggedright\arraybackslash}X >{\raggedright\arraybackslash}X >{\raggedright\arraybackslash}X@{}}
\toprule
\textbf{Dimension} & \textbf{What It Binds} & \textbf{Who Controls} & \textbf{Verification} \\
\midrule
Developer Identity &
  Developer's Ed25519 signing key &
  Developer (build time) &
  GAR verifies developer enrollment \\[3pt]
Code Package &
  Cryptographic digest of the agent's code or container image &
  Build pipeline &
  GAR checks code hash at issuance \\[3pt]
Operator Identity &
  OIDC-verified human operator (e.g., via Google, GitHub SSO) &
  Human operator (provisioning) &
  GAR binds OIDC token to AIC \\[3pt]
Operational Context &
  Tenant ID, deployment environment, region &
  Platform operator &
  GAR records at issuance; inherited by delegates \\
\bottomrule
\end{tabularx}
\end{table}

This four-dimensional binding addresses threat T1 (identity spoofing) from
\Cref{tab:threats}: an attacker cannot forge an AIC without simultaneously
controlling the developer's signing key, producing a matching code digest,
presenting valid OIDC credentials, and registering with the correct
operational context.

\parab{Contrast with existing approaches}
DID-based identity~\cite{did-vc-agents,saavedra-delegation} binds a public
key to a decentralized identifier but does not natively encode developer,
code, or operator dimensions. SPIFFE~\cite{spiffe-agents} provides workload
identity attestation via platform-specific mechanisms but does not model
capability boundaries or delegation semantics. The AIP
protocol~\cite{aip-protocol} introduces Invocation-Bound Capability Tokens
without separating developer and operator dimensions in the credential model. \sys's four-dimensional binding provides richer provenance than
workload-centric models such as SPIFFE/SVID (used by
AgentMesh~\cite{agentmesh}) and, among the identity proposals surveyed in
\S\ref{sec:related}, is the only one we found that jointly binds developer,
code, operator, and operational context within a single verifiable credential.

\subsection{Global Agent Registry (GAR)}
\label{subsec:gar}

The GAR serves as the trust anchor for the \sys ecosystem, playing a role
analogous to a Certificate Authority in traditional PKI. Its responsibilities
include:

\begin{enumerate}[leftmargin=2em,itemsep=2pt]
  \item \textbf{Developer enrollment:} onboarding developers with verified
    signing keys and policy governance.
  \item \textbf{AIC issuance:} validating that the requested capability
    boundary $\Smax$ is a subset of the developer's declared maximum,
    verifying the developer signature, binding the OIDC operator identity,
    and signing the AIC with the GAR root key.
  \item \textbf{AIC lookup and verification:} enabling any party to resolve
    an agent's AIC by identifier and verify its signature chain.
  \item \textbf{Revocation:} maintaining a revocation registry; revoking an
    AIC atomically invalidates it and all its delegated descendants
    (cascading revocation, detailed in \S\ref{subsec:delegation}).
\end{enumerate}

The GAR can be deployed as a centralized service (suitable for enterprise
environments), a federated constellation (analogous to federated CAs), or
backed by a distributed ledger for environments that require decentralized
trust anchoring. This flexibility is a direct consequence of Principle~P3.

\subsection{Key Protection Tiers}
\label{subsec:key-protection}

The agent's private key $\Kpriv$ is the root of its cryptographic identity.
\sys defines three protection tiers, allowing deployments to choose the
appropriate security-cost trade-off:

\begin{table}[h]
\centering
\small
\begin{tabularx}{\textwidth}{@{}c l X l@{}}
\toprule
\textbf{Tier} & \textbf{Mechanism} & \textbf{Security Level} & \textbf{Deployment} \\
\midrule
1 & OS file isolation (\texttt{0o600}) & Moderate & Dev / testing \\
2 & OS keychain (Keychain, DPAPI) & Good (HW-backed) & Production \\
3 & TEE enclave (SGX, Nitro) & Strong (key never leaves HW) & High-assurance \\
\bottomrule
\end{tabularx}
\caption{Key protection tiers. $\Kpriv$ never leaves the local trust boundary regardless of tier.}
\label{tab:key-tiers}
\end{table}

Regardless of the tier, a fundamental invariant holds: $\Kpriv$ \emph{never
leaves the isolated trust boundary}. All signing operations are mediated by the
agent's designated kernel, which acts as an HSM-like custodian. External agents and
even the agent's own application logic interact only with opaque
\emph{handles}---they can request signatures but never access the raw key
material.

\subsection{Identity Lifecycle}
\label{subsec:identity-lifecycle}

An AIC progresses through a well-defined lifecycle:

\begin{enumerate}[leftmargin=2em,itemsep=2pt]
  \item \textbf{Provisioning:} The operator submits a registration request
    (agent ID, capabilities, OIDC token) to the kernel. The kernel generates
    a fresh Ed25519 keypair, constructs the AIC payload, and forwards it to
    the GAR for signing. The signed AIC is returned and cached locally.
  \item \textbf{Active use:} The AIC is used for mutual attestation
    (\S\ref{sec:authentication}), session token issuance, and delegation.
  \item \textbf{Rotation:} Key rotation generates a new keypair, revokes the
    old AIC, and issues a replacement preserving all identity dimensions.
    Active sessions are invalidated to prevent stale credential use.
  \item \textbf{Revocation:} An AIC can be revoked by the operator, the GAR
    administrator, or automatically upon expiration. Revocation cascades to
    all delegated descendants (\S\ref{subsec:delegation}), ensuring that
    revoking a parent instantly invalidates the entire subtree.
\end{enumerate}

The lifecycle is designed so that identity is \emph{persistent} (the agent's
identity dimensions survive key rotation) while credentials are
\emph{ephemeral} (AICs have bounded validity and can be revoked at any time).
This distinction is crucial for long-lived agents that operate across
sessions and deployments.

%% file: figures/aic-structure.tex
\begin{tikzpicture}[
  dim/.style={
    draw=diborder, rounded corners=3pt, minimum width=3cm,
    minimum height=1.8cm, font=\sffamily\small, align=center,
    fill=white, text=didarkgray
  },
  aic/.style={
    draw=diblue, line width=1.2pt, rounded corners=6pt,
    minimum width=5cm, minimum height=2.2cm, font=\sffamily,
    fill=diblue!5, align=center, text=diblue
  },
  gar/.style={
    draw=dipurple, line width=1pt, rounded corners=4pt,
    minimum width=3.5cm, minimum height=1.2cm, font=\sffamily\small,
    fill=dipurple!8, text=dipurple, align=center
  },
  arr/.style={-{Stealth[length=5pt]}, thick, color=didarkgray},
  lbl/.style={font=\sffamily\scriptsize, text=gray},
]

\node[aic] (aic) at (0,0) {
  \textbf{Agent Identity Card (AIC)}\\[4pt]
  $\Sign_{\GAR}\!\bigl(\DIDag \;\|\; \Kpub \;\|\; \Smax\bigr)$
};

\node[dim, fill=diblue!8] (dev) at (-5, 2.5)
  {\textbf{Developer Identity}\\[2pt]
   Ed25519 signing key\\
   {\scriptsize controlled at build time}};

\node[dim, fill=diaccent!8] (code) at (5, 2.5)
  {\textbf{Code Package}\\[2pt]
   SHA-256 digest\\
   {\scriptsize set by build pipeline}};

\node[dim, fill=digreen!8] (op) at (-5, -2.5)
  {\textbf{Operator Identity}\\[2pt]
   OIDC-verified human\\
   {\scriptsize bound at provisioning}};

\node[dim, fill=diorange!8] (ctx) at (5, -2.5)
  {\textbf{Operational Context}\\[2pt]
   Tenant, environment\\
   {\scriptsize set by platform}};

\draw[arr] (dev.south east) -- (aic.north west);
\draw[arr] (code.south west) -- (aic.north east);
\draw[arr] (op.north east) -- (aic.south west);
\draw[arr] (ctx.north west) -- (aic.south east);

\node[gar] (gar) at (0, -4.5)
  {\textbf{Global Agent Registry}\\
   signs with GAR root key};
\draw[arr, dipurple] (gar.north) -- (aic.south)
  node[midway, right=3pt, lbl] {$\Sign_{\GAR}$};

\end{tikzpicture}

%% file: sections/registration.tex
\section{Layer~1: Registration, Discovery \& Semantic Interoperability}
\label{sec:registration}

Layer~0 establishes \emph{who an agent is}; Layer~1 establishes \emph{what
an agent can do and how to find it}. This layer bridges identity to
interaction by providing the infrastructure for agents to advertise their
capabilities, discover peers, and cryptographically verify the authenticity
of advertised skills before any trust negotiation begins.

The central insight of Layer~1 is that discovery must be a \emph{security
primitive}: every capability claim an agent advertises is backed by a signed
Verifiable Credential (VC) bound to the agent's DID and verifiable against
the AIC trust chain. This transforms agent discovery from a directory
lookup into a trust-establishing protocol step.

\subsection{Capability-Aware Registration}
\label{subsec:registration}

When an agent is provisioned (Layer~0), its AIC already contains a capability
boundary~$\Smax$. Layer~1 extends this with a richer \emph{registration
record} that the agent publishes to the GAR or a federated discovery service:

\begin{protocolbox}[Registration Record]
\small
\begin{tabbing}
\hspace{1.5em}\= \hspace{7em}\= \kill
\> \textbf{agent\_id}     \> $\DIDag$ from the AIC \\
\> \textbf{aic\_ref}      \> reference to the signed AIC \\
\> \textbf{capabilities}  \> structured list from the capability vocabulary \\
\> \textbf{semantic\_tags} \> tags for capability-aware search \\
\> \textbf{endpoints}     \> transport endpoints (HTTP, gRPC, WebSocket) \\
\> \textbf{protocols}     \> supported protocols (A2A, MCP, ANP) \\
\> \textbf{manifests}     \> list of signed skill/tool manifest VCs \\
\> \textbf{metadata}      \> description, version, pricing hints
\end{tabbing}
\end{protocolbox}

The registration record is itself signed by the agent's $\Kpriv$ (via the
kernel), ensuring that only the agent can create or update its own record.
The GAR or discovery service verifies this signature against the AIC's
$\Kpub$ before accepting the registration.

\subsection{Semantic Capability Tagging}
\label{subsec:semantic-tags}

Raw capability enumerations (e.g., \texttt{fs.read}, \texttt{network.fetch})
describe what system surfaces an agent can touch but not the \emph{semantic
domain} of the agent's expertise. \sys introduces a two-level tagging scheme:

\begin{enumerate}[leftmargin=2em,itemsep=2pt]
  \item \textbf{System capabilities} (from Layer~0): a closed vocabulary of
    permission classes bound to the AIC. These are machine-enforced at
    runtime.
  \item \textbf{Semantic tags} (from Layer~1): an open vocabulary of
    domain-specific labels (e.g., \texttt{legal-review},
    \texttt{code-generation}, \texttt{financial-analysis}) that describe the
    agent's expertise at a human-understandable and LLM-parseable level.
\end{enumerate}

Semantic tags are \emph{not} capability grants---they carry no enforcement
weight. Their purpose is to enable capability-aware discovery: a requester
searching for a ``code review'' agent can filter candidates by semantic tag,
then verify actual capabilities via the AIC and manifest VCs before
Layer~2 negotiation.
This separation prevents semantic labels from being abused as implicit
permission escalation vectors.

\subsection{Capability-Aware Discovery}
\label{subsec:discovery}

\sys treats discovery as a security primitive rather than a directory lookup.
A discovery response is only useful if the requester can verify not just that
an agent claims a skill or tool, but that the advertised capability is backed by a
signed manifest VC and fits within the agent's AIC capability boundary. This
turns Layer~1 into \emph{capability-aware discovery}: semantic search narrows
the candidate set, while DID-bound manifest VCs and AIC-bound capabilities
make the result safe to consume during Layer~2 negotiation.

\sys supports two discovery modes that can coexist within the same ecosystem:

\parab{Registry-mediated discovery}
Agents query the GAR (or a federated registry constellation) with structured
queries over capabilities, semantic tags, protocol support, and trust
attributes (e.g., minimum identity assurance level, trusted developer list).
The registry returns matching registration records, each verifiable against
the corresponding AIC. This mode is analogous to DNS resolution and is
suitable for enterprise and platform-managed environments.

\parab{Peer-to-peer discovery}
In decentralized environments, agents can discover peers via protocol-native
mechanisms (e.g., ANP's DID-based discovery~\cite{anp-whitepaper}, A2A's
well-known Agent Cards~\cite{a2a-protocol}). \sys does not replace these
mechanisms but \emph{augments} them: once a candidate agent is discovered via
any means, its AIC can be resolved and verified through the GAR, adding a
trust verification step that the native discovery protocol may lack.

Both modes converge on the same verification flow: the discovered agent's AIC
is verified against the GAR's trust chain, and each manifest VC in its
registration record is validated via the four-check protocol
(\S\ref{subsec:manifests}). Discovery thus answers three questions
simultaneously: ``who is out there'' (AIC verification), ``what can they
do'' (manifest VC verification), and ``should I trust them'' (subject
binding + permission alignment).

\subsection{Verifiable Skill \& Tool Manifests}
\label{subsec:manifests}

An agent's capabilities are realized through two mechanisms:
\begin{itemize}[leftmargin=2em,itemsep=2pt]
  \item \textbf{Skills}---reusable capability modules (code packages) that
    run within the agent's execution environment.
  \item \textbf{Tools}---external service bindings (API endpoints, MCP
    servers) that the agent invokes over the network.
\end{itemize}

\noindent In \sys, both skills and tools are represented as \emph{signed
Verifiable Credentials (VCs)} cryptographically bound to the agent's DID
and verifiable against the AIC trust chain. This design ensures that
capability claims---whether local code or remote services---carry the same
structural trust guarantees as agent identity itself. The two types share a
common credential structure but differ in their issuance model, reflecting
their distinct trust relationships.

\parab{Credential structure}
Both skill and tool manifest VCs follow the W3C Verifiable Credentials
data model~\cite{w3c-vc} adapted for agent semantics:

\begin{protocolbox}[Skill Manifest VC]
\small
\begin{tabbing}
\hspace{1.5em}\= \hspace{11em}\= \kill
\> \textbf{@context}           \> W3C VC context + \sys skill vocabulary \\
\> \textbf{id}                 \> unique credential URI \\
\> \textbf{type}               \> [``VerifiableCredential'', ``SkillManifest''] \\
\> \textbf{issuer}             \> distributor DID (skill author) \\
\> \textbf{credentialSubject}  \> \{ \\
\> \quad\textbf{id}               \> $\DIDag$ (the agent holding this skill) \\
\> \quad\textbf{manifest\_id}     \> unique manifest identifier \\
\> \quad\textbf{perms\_required}  \> system capabilities the skill needs \\
\> \quad\textbf{caps\_provided}   \> what the skill enables for the agent \\
\> \quad\textbf{code\_hash}       \> digest of the skill code package \\
\> \textbf{\}}                 \> \\
\> \textbf{proof}              \> Ed25519 signature by the distributor \\
\> \textbf{gar\_endorsement}   \> optional GAR endorsement
\end{tabbing}
\end{protocolbox}

\begin{protocolbox}[Tool Manifest VC]
\small
\begin{tabbing}
\hspace{1.5em}\= \hspace{11em}\= \kill
\> \textbf{@context}           \> W3C VC context + \sys tool vocabulary \\
\> \textbf{id}                 \> unique credential URI \\
\> \textbf{type}               \> [``VerifiableCredential'', ``ToolManifest''] \\
\> \textbf{issuer}             \> tool provider DID (service operator) \\
\> \textbf{credentialSubject}  \> \{ \\
\> \quad\textbf{id}               \> $\DIDag$ (the agent authorized to use this tool) \\
\> \quad\textbf{manifest\_id}     \> unique manifest identifier \\
\> \quad\textbf{perms\_required}  \> capabilities the tool invocation needs \\
\> \quad\textbf{caps\_provided}   \> what the tool enables for the agent \\
\> \quad\textbf{endpoint}         \> service URI (MCP server, REST API, etc.) \\
\> \quad\textbf{api\_hash}        \> digest of the tool's interface schema \\
\> \textbf{\}}                 \> \\
\> \textbf{proof}              \> Ed25519 signature by the tool provider \\
\> \textbf{gar\_endorsement}   \> optional GAR endorsement
\end{tabbing}
\end{protocolbox}

In both cases the \textbf{credentialSubject.id} field binds the manifest to
a specific agent DID ($\DIDag$). This binding means that a manifest VC is
non-transferable: even if an attacker obtains the credential, it cannot be
presented on behalf of a different agent because the verifier checks that
the subject DID matches the presenter's AIC.

\parab{Skill manifest issuance}
The lifecycle of a \emph{skill} manifest VC proceeds as follows:
\begin{enumerate}[leftmargin=2em,itemsep=2pt]
  \item \textbf{Distributor publishes skill.} The skill author (the
    distributor) creates the manifest, signs it with their distributor key,
    and optionally submits it to the GAR for endorsement.
  \item \textbf{Agent acquires skill.} When an agent installs a skill, the
    kernel verifies the distributor's signature and (if present) the GAR
    endorsement. The kernel then issues a \emph{binding proof}: a secondary
    signature using the agent's $\Kpriv$ that attests ``I hold this skill
    and my AIC permits its required capabilities.''
  \item \textbf{Agent presents during discovery.} The agent includes its
    bound skill manifest VCs in its registration record.
\end{enumerate}

\parab{Tool manifest issuance}
Tools differ from skills because their trust anchor is the \emph{tool
provider} (the external service operator), not a code distributor:
\begin{enumerate}[leftmargin=2em,itemsep=2pt]
  \item \textbf{Tool provider registers service.} The service operator
    publishes a tool manifest describing the tool's endpoint, interface
    schema, and required permissions. The provider signs the manifest with
    their provider key and optionally obtains a GAR endorsement.
  \item \textbf{Provider issues VC to agent.} When an agent requests access
    to a tool, the provider verifies the agent's AIC and checks that the
    agent's capability boundary $\Smax$ is a superset of the tool's
    \texttt{perms\_required}. If satisfied, the provider issues a tool
    manifest VC with \texttt{credentialSubject.id} set to the agent's
    $\DIDag$---effectively granting the agent verifiable authorization to
    invoke the service.
  \item \textbf{Agent presents during discovery.} The agent includes its
    tool manifest VCs alongside skill manifest VCs in its registration
    record. Requesters can now verify that the agent not only claims access
    to a tool but has been explicitly authorized by the tool provider.
\end{enumerate}

\noindent In both flows, the agent's registration record ultimately
contains a set of manifest VCs---some for skills, some for tools---each
independently verifiable by any requester or registry.

\parab{Verification during discovery}
When a requester discovers an agent and retrieves its registration record,
the following checks are performed on each manifest VC before the agent is
considered a valid candidate for Layer~2 negotiation:
\begin{itemize}[leftmargin=2em,itemsep=2pt]
  \item \textbf{Supply-chain verification:} The issuer's signature
    (skill distributor or tool provider) and any GAR endorsement are
    verified, preventing supply-chain attacks analogous to the ``ClawHavoc''
    incident~\cite{identity-aware-governance}.
  \item \textbf{Subject binding:} The \texttt{credentialSubject.id} must
    match the discovered agent's $\DIDag$. This prevents manifest replay
    attacks where an adversary copies another agent's manifest credentials.
  \item \textbf{Permission alignment:} The manifest's
    \texttt{perms\_required} must be a subset of the agent's AIC capability
    boundary~$\Smax$. If a skill or tool requires permissions beyond the agent's
    boundary, the manifest is rejected, ensuring that discovery cannot
    surface agents whose advertised claims exceed their actual authorization.
  \item \textbf{Freshness:} The manifest VC must not be expired or revoked
    (checked against the GAR's revocation list or status endpoint).
\end{itemize}

This four-check verification protocol transforms discovery from a trust-me
directory into a \emph{verify-then-interact} security boundary. The
consequence is structural: an agent cannot advertise capabilities it does
not possess, because every advertised skill or tool must be backed by a VC whose
subject binding, issuer provenance, and permission alignment are all
independently verifiable.

%% file: sections/authentication.tex
\section{Layer~2: Trust Negotiation}
\label{sec:authentication}

Layer~2 is the interaction layer of \sys. It governs three critical
processes: \emph{mutual attestation} (how two agents establish trust),
\emph{delegation} (how a parent agent creates bounded child identities), and
\emph{access control} (how a responder decides what a requester may do).
Together, these processes address threats T2--T4 from \Cref{tab:threats}.

\begin{protocolbox}[]
\subsection{Mutual Attestation Protocol}
\label{subsec:attestation}

When two agents wish to interact, neither should blindly trust the other. The
mutual attestation protocol establishes bilateral trust through a
challenge-response exchange grounded in the agents' AICs.

\smallskip
\small
We write \KerA and \KerB for the kernels that hold $\Kpriv^A$ and
$\Kpriv^B$ respectively. When both agents reside on the same host,
\KerA${} = {}$\KerB and the protocol reduces to local operations.
\begin{center}
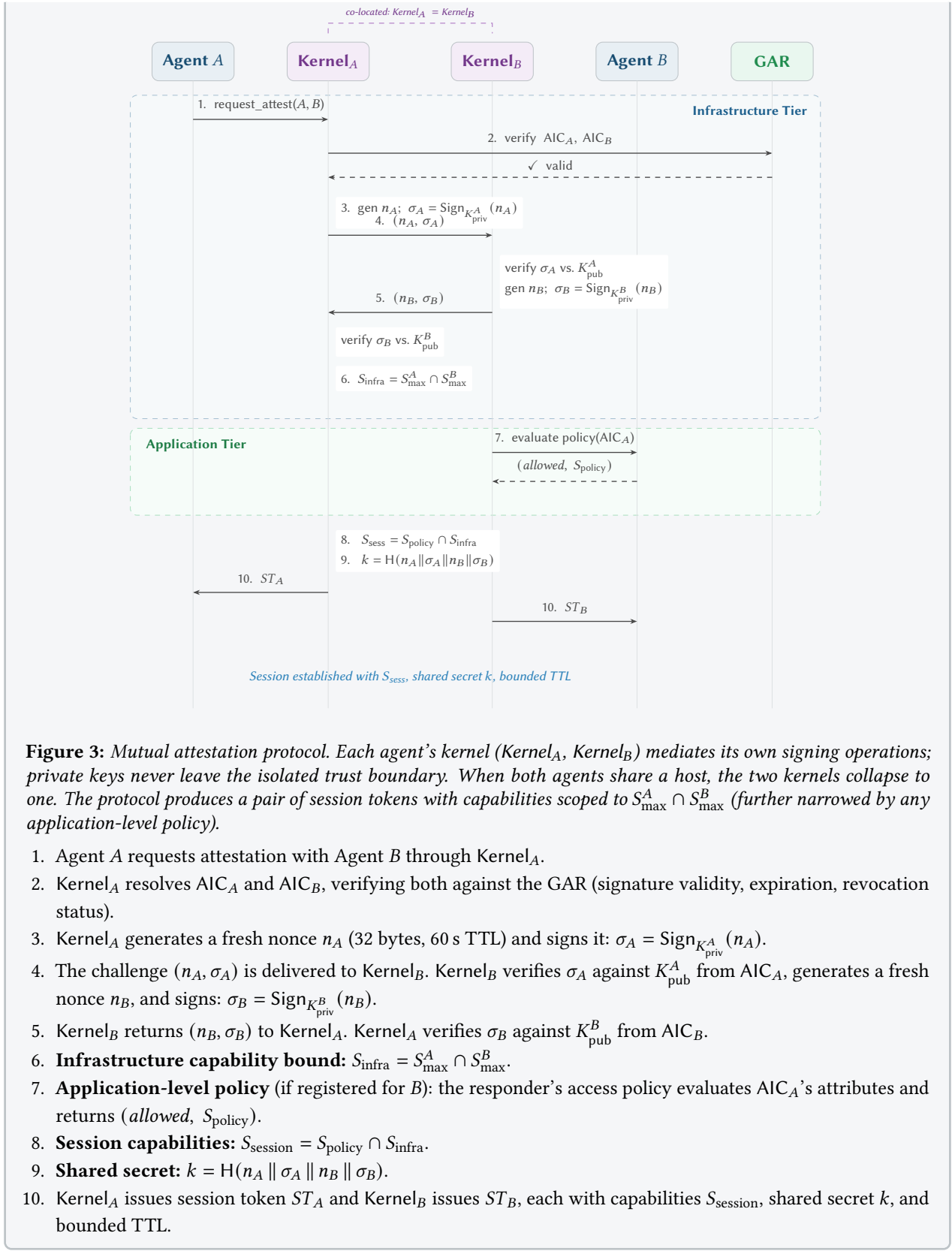

  \scalebox{0.85}{\input{figures/a2a-attestation}}
\end{center}
\vspace{0.15\baselineskip}
\captionof{figure}{Mutual attestation protocol. Each agent's kernel (\KerA, \KerB)
mediates its own signing operations; private keys never leave the isolated
trust boundary. When both agents share a host, the two kernels collapse to
one. The protocol produces a pair of session tokens with capabilities
scoped to $\Smax^A \cap \Smax^B$ (further narrowed by any
application-level policy).}
\label{fig:attestation}
\vspace{0.2\baselineskip}
\begin{enumerate}[leftmargin=1.5em,itemsep=0pt]
  \item Agent~$A$ requests attestation with Agent~$B$ through \KerA.
  \item \KerA resolves $\AIC_A$ and $\AIC_B$, verifying both against
    the GAR (signature validity, expiration, revocation status).
  \item \KerA generates a fresh nonce $n_A$ (32 bytes, 60\,s TTL) and
    signs it: $\sigma_A = \Sign_{\Kpriv^A}(n_A)$.
  \item The challenge $(n_A, \sigma_A)$ is delivered to \KerB.
    \KerB verifies $\sigma_A$ against $\Kpub^A$ from $\AIC_A$,
    generates a fresh nonce~$n_B$, and signs:
    $\sigma_B = \Sign_{\Kpriv^B}(n_B)$.
  \item \KerB returns $(n_B, \sigma_B)$ to \KerA.
    \KerA verifies $\sigma_B$ against $\Kpub^B$ from $\AIC_B$.
  \item \textbf{Infrastructure capability bound:}
    $S_{\text{infra}} = \Smax^A \cap \Smax^B$.
  \item \textbf{Application-level policy} (if registered for $B$): the
    responder's access policy evaluates $\AIC_A$'s attributes and returns
    $({\mathit{allowed}},\; S_{\text{policy}})$.
  \item \textbf{Session capabilities:}
    $S_{\text{session}} = S_{\text{policy}} \cap S_{\text{infra}}$.
  \item \textbf{Shared secret:}
    $k = \Hash(n_A \,\|\, \sigma_A \,\|\, n_B \,\|\, \sigma_B)$.
  \item \KerA issues session token $\mathit{ST}_A$ and \KerB issues
    $\mathit{ST}_B$, each with capabilities $S_{\text{session}}$,
    shared secret~$k$, and bounded TTL.
\end{enumerate}
\end{protocolbox}

\begin{remark}[Deployment Topologies]
\label{rem:attestation-topologies}
The protocol above is parameterized by the relationship between \KerA
and \KerB. Three instantiations cover the practical deployment spectrum:

\begin{description}[leftmargin=1.5em,itemsep=3pt]
  \item[Co-located (single kernel).]
    When $A$ and $B$ reside on the same host, \KerA${} = {}$\KerB and
    steps~3--5 collapse to kernel-internal operations with no network
    round-trip. This is the natural mode for \deepkernel-managed
    multi-agent orchestration on a single device.

  \item[Remote-mediated (Trusted Attestation Service).]
    A standalone service---architecturally identical to an agent kernel
    but deployed as a remote endpoint---coordinates the exchange. Each
    agent's designated kernel retains $\Kpriv$ and signs within its
    trust boundary; the \TAS
    relays challenges, caches GAR verification results, and issues session
    tokens. The \TAS plays a role analogous to an OIDC Provider in human
    web SSO: it centralizes session establishment while each agent
    retains sovereign control of its private key.

  \item[Direct peer-to-peer.]
    Each kernel signs its own nonce and verifies the counterpart's signature
    independently via the GAR. No trusted intermediary is required.
    This mode is fully decentralized but incurs higher latency: each side
    performs an independent GAR round-trip, and every attestation requires
    a full network challenge-response exchange. AIC caching with
    revocation-push notifications (\S\ref{sec:discussion}) can amortize
    the GAR cost.
\end{description}

The three modes can coexist within a single ecosystem: co-located
attestation for intra-device workflows, mediated attestation for
organizational clusters, and direct P2P for open cross-organization
interactions.
\end{remark}

\parab{Anti-spoofing guarantees}
Session tokens are bound to specific agent instances (AIC + execution
context). Tokens expire on agent termination; replay is impossible because
nonces are single-use and time-bounded. Private keys never leave the
agent's isolated kernel, so stolen tokens are useless outside the authorized
execution context.

\subsection{AIC Delegation Chain}
\label{subsec:delegation}

In multi-agent orchestration scenarios, a parent agent often needs to
dynamically spawn task-specific child agents. \sys supports this through
\emph{AIC delegation}: the parent requests the kernel to issue a child AIC
with strictly attenuated capabilities, forming a cryptographic chain
analogous to an X.509 intermediate-CA hierarchy.

\begin{protocolbox}[Delegation Protocol]
\small
\begin{enumerate}[leftmargin=1.5em,itemsep=1pt]
  \item Parent requests child AIC from the kernel, specifying desired
    capabilities $S_{\text{child}} \subseteq \Smax^{\text{parent}}$.
  \item Kernel verifies the parent's AIC chain (iterative walk to GAR root).
  \item Kernel checks:
    $S_{\text{child}} \subseteq \Smax^{\text{parent}}$
    (no capability escalation).
  \item Kernel checks: delegation depth $\leq$ configured ceiling.
  \item Kernel generates fresh Ed25519 keypair for the child.
  \item GAR issues child AIC signed by the parent's key, with:
    \begin{itemize}[leftmargin=1.5em,itemsep=0pt]
      \item $\Smax^{\text{child}} = S_{\text{child}}$
      \item Inherited \texttt{tenant\_id} and
        \texttt{deployment\_environment}
      \item Validity capped to parent's remaining lifetime
    \end{itemize}
  \item Child AIC is returned; parent-to-child link is recorded.
\end{enumerate}
\vspace{0.35\baselineskip}
\begin{center}
  \input{figures/capability-chain}
\end{center}
\vspace{0.25\baselineskip}
\captionof{figure}{Monotonic capability attenuation chain. Each party can only
\emph{narrow} the boundary set by the preceding party. The runtime
intersection further constrains per-invocation permissions.}
\label{fig:capability-chain}
\end{protocolbox}

The delegation mechanism provides six structural security properties:

\begin{table}[h]
\centering
\small
\caption{Security properties of AIC delegation.}
\label{tab:delegation-props}
\begin{tabularx}{\textwidth}{@{}l >{\raggedright\arraybackslash}X@{}}
\toprule
\textbf{Property} & \textbf{Mechanism} \\
\midrule
Monotonic attenuation &
  $\Smax^{\text{child}} \subseteq \Smax^{\text{parent}}$ enforced
  cryptographically at delegation time. \\
Tenant isolation &
  Child inherits \texttt{tenant\_id} from parent; cross-tenant delegation is
  structurally impossible. \\
Depth bounding &
  $d_{\text{eff}} = \min(d_{\text{kernel}},\, d_{\text{request}})$; the
  request can only tighten, never loosen. \\
Cascading revocation &
  Revoking a parent AIC atomically revokes all descendants; chain
  verification fails for the entire subtree. \\
Validity capping &
  Child's \texttt{expires\_at} $\leq$ parent's \texttt{expires\_at};
  parent expiry cascades temporally. \\
Cycle resistance &
  Iterative chain verification with visited set and hard hop cap; no
  recursion overflow or cycle attacks. \\
\bottomrule
\end{tabularx}
\end{table}

\subsection{Two-Tier A2A Access Control}
\label{subsec:access-control}

A pure capability-intersection model (\S\ref{subsec:attestation}, step~6) is
necessary but insufficient: the responder agent may wish to restrict
interactions based on attributes beyond raw capabilities---for example,
allowing only agents from trusted developers, requiring a minimum identity
assurance level, or enforcing same-tenant isolation.

\sys introduces a \emph{two-tier access control model} that separates
platform-enforced trust from application-level policy:

\begin{enumerate}[leftmargin=2em]
  \item \textbf{Infrastructure tier (mandatory, kernel-enforced):}
    Cryptographic AIC verification via the GAR. This tier answers: ``Is this
    agent who it claims to be?'' Both parties' AICs are verified for
    signature validity, expiration, and revocation. The infrastructure
    capability bound $S_{\text{infra}} = \Smax^A \cap \Smax^B$ is computed.

  \item \textbf{Application tier (optional, agent-defined):}
    The responder agent registers a declarative \emph{access policy} that
    evaluates the requester's AIC attributes:
    \begin{itemize}[leftmargin=1.5em,itemsep=1pt]
      \item Required capabilities (requester must possess specific caps)
      \item Trusted developers (glob patterns on \texttt{developer\_id})
      \item Trusted operators (glob patterns on \texttt{operator\_id})
      \item Minimum identity assurance level (IAL threshold)
      \item Maximum risk level
      \item Same-tenant requirement
    \end{itemize}
    The application tier answers: ``Does this specific responder want to work
    with this specific requester?'' It produces a (possibly narrowed) set of
    \emph{granted capabilities} that is intersected with the infrastructure
    bound.
\end{enumerate}

The final session capability set is:
\begin{equation}
\label{eq:session-caps}
S_{\text{session}} = S_{\text{policy}} \cap \bigl(\Smax^A \cap \Smax^B\bigr)
\end{equation}

This two-tier design has three key advantages over single-layer approaches:
\begin{itemize}[leftmargin=2em,itemsep=2pt]
  \item \textbf{Composability:} Infrastructure and application concerns are
    independently auditable and evolvable.
  \item \textbf{Backward compatibility:} Agents without a registered policy
    fall back to the infrastructure tier alone.
  \item \textbf{Fine-grained control:} The responder can make admission
    decisions based on identity attributes that raw capability intersection
    cannot express.
\end{itemize}

\subsection{The Capability Narrowing Chain}
\label{subsec:narrowing-chain}

\sys's approach to authorization can be summarized as a four-stage
\emph{narrowing chain} in which each stage can only tighten the permission
boundary:

\begin{center}
\small
\begin{tabular}{@{}r@{\;\;}c@{\;\;}l@{\;\;}l@{}}
\textbf{Stage 1:} & Developer & declares & $S_{\text{declared}}$ \\
\textbf{Stage 2:} & Operator  & selects  & $\Smax \subseteq S_{\text{declared}}$ \\
\textbf{Stage 3:} & GAR       & enforces \& signs & $\Smax$ (subset-checked, signed AIC) \\
\textbf{Stage 4:} & Runtime   & intersects & $S_{\text{session}} \subseteq \Smax^A \cap \Smax^B$ \\
\end{tabular}
\end{center}

At no point can any party \emph{widen} the boundary set by the preceding
party. This is enforced cryptographically: the developer's signature covers
$S_{\text{declared}}$; the GAR's signature covers $\Smax \subseteq
S_{\text{declared}}$; the session token is kernel-signed over
$S_{\text{session}} \subseteq \Smax^A \cap \Smax^B$. Any attempt to insert
additional capabilities requires forging a signature.

This monotonic attenuation provides \emph{structural least-privilege}: the
property holds by construction, not by correct policy configuration.
Misconfiguring a policy can deny legitimate access but can never grant
unauthorized capabilities.

%% file: figures/a2a-attestation.tex
%
%
\begin{tikzpicture}[
  entity/.style={
    draw=diborder, rounded corners=4pt, minimum width=1.7cm,
    minimum height=0.8cm, font=\sffamily\small\bfseries,
    fill=#1!10, text=#1!80!black
  },
  msg/.style={-{Stealth[length=4pt]}, semithick, color=didarkgray},
  ret/.style={-{Stealth[length=4pt]}, semithick, color=didarkgray, dashed},
  msglbl/.style={font=\sffamily\scriptsize, text=didarkgray, midway, above=1pt},
  note/.style={
    font=\sffamily\scriptsize, text=didarkgray, align=left,
    anchor=north west,
    inner sep=3pt, fill=white, rounded corners=1pt,
  },
  tierlbl/.style={
    font=\sffamily\scriptsize\bfseries, anchor=north west,
  },
]

\def\xA{0}       
\def\xKA{2.8}    
\def\xKB{6.2}    
\def\xB{9.2}     
\def\xG{12.0}    

\node[entity=diblue]   (agA)  at (\xA, 0)   {Agent $A$};
\node[entity=dipurple] (kerA) at (\xKA, 0)  {\KerA};
\node[entity=dipurple] (kerB) at (\xKB, 0)  {\KerB};
\node[entity=diblue]   (agB)  at (\xB, 0)   {Agent $B$};
\node[entity=digreen]  (gar)  at (\xG, 0)   {GAR};

\draw[dipurple!60, dashed, semithick]
  ([yshift=5pt]kerA.north) -- ([yshift=11pt]kerA.north)
  -- ([yshift=11pt]kerB.north) -- ([yshift=5pt]kerB.north);
\node[font=\sffamily\tiny\itshape, text=dipurple!80!black, anchor=south]
  at ($(kerA.north)!0.5!(kerB.north) + (0, 11pt)$)
  {co-located: \KerA${} = {}$\KerB};

\def\lifebot{-13.4}
\foreach \x in {\xA, \xKA, \xKB, \xB, \xG} {
  \draw[diborder!50, thin] (\x, -0.4) -- (\x, \lifebot);
}

\begin{scope}[on background layer]
  \fill[diblue!4, rounded corners=4pt]
    (-1.3, -0.7) rectangle (13.0, -7.4);
  \draw[diblue!40, dashed, rounded corners=4pt]
    (-1.3, -0.7) rectangle (13.0, -7.4);
\end{scope}
\node[tierlbl, text=diblue, anchor=north east] at (12.85, -0.78) {Infrastructure Tier};

\draw[msg] (\xA, -1.2) -- (\xKA, -1.2)
  node[msglbl] {\textsf{1.}\; request\_attest($A$,\,$B$)};

\draw[msg] (\xKA, -1.9) -- (\xG, -1.9)
  node[msglbl] {\textsf{2.}\; verify\; $\AIC_A$,\, $\AIC_B$};
\draw[ret] (\xG, -2.4) -- (\xKA, -2.4)
  node[msglbl] {\checkmark\; valid};

\node[note] (opsA) at (\xKA+0.15, -2.8) {%
  \textsf{3.}\; gen $n_A$;\; $\sigma_A = \Sign_{\Kpriv^A}(n_A)$%
};

\draw[msg] (\xKA, -3.6) -- (\xKB, -3.6)
  node[msglbl] {\textsf{4.}\; $(n_A,\, \sigma_A)$};

\node[note] (opsB) at (\xKB+0.15, -4.0) {%
  verify $\sigma_A$ vs.\ $\Kpub^A$\\[1pt]
  gen $n_B$;\; $\sigma_B = \Sign_{\Kpriv^B}(n_B)$%
};

\draw[msg] (\xKB, -5.2) -- (\xKA, -5.2)
  node[msglbl] {\textsf{5.}\; $(n_B,\, \sigma_B)$};

\node[note] (verA) at (\xKA+0.15, -5.5) {%
  verify $\sigma_B$ vs.\ $\Kpub^B$%
};

\node[note] (infra) at (\xKA+0.15, -6.3) {%
  \textsf{6.}\; $S_{\text{infra}} = \Smax^A \cap \Smax^B$%
};

\begin{scope}[on background layer]
  \fill[digreen!4, rounded corners=4pt]
    (-1.3, -7.6) rectangle (13.0, -9.4);
  \draw[digreen!40, dashed, rounded corners=4pt]
    (-1.3, -7.6) rectangle (13.0, -9.4);
\end{scope}
\node[tierlbl, text=digreen!70!black] at (-1.1, -7.7) {Application Tier};

\draw[msg] (\xKB, -8.1) -- (\xB, -8.1)
  node[msglbl] {\textsf{7.}\; evaluate policy($\AIC_A$)};
\draw[ret] (\xB, -8.7) -- (\xKB, -8.7)
  node[msglbl] {$(\mathit{allowed},\; S_{\text{policy}})$};


\node[note] (sess) at (\xKA+0.15, -9.7) {%
  \textsf{8.}\;\; $S_{\text{sess}} = S_{\text{policy}} \cap S_{\text{infra}}$\\[1pt]
  \textsf{9.}\;\; $k = \Hash(n_A \| \sigma_A \| n_B \| \sigma_B)$%
};

\draw[msg] (\xKA, -11.0) -- (\xA, -11.0)
  node[msglbl] {\textsf{10.}\; $\mathit{ST}_A$};
\draw[msg] (\xKB, -11.6) -- (\xB, -11.6)
  node[msglbl] {\textsf{10.}\; $\mathit{ST}_B$};

\node[font=\sffamily\scriptsize\itshape, text=diaccent, anchor=north]
  at (0.5*\xKA+0.5*\xKB, -12.5)
  {Session established with $S_{\text{sess}}$, shared secret $k$, bounded TTL};

\end{tikzpicture}

%% file: figures/capability-chain.tex
\begin{tikzpicture}[
  stage/.style={
    draw=diborder, rounded corners=4pt, minimum width=2.8cm,
    minimum height=1.6cm, font=\sffamily\small, align=center,
    fill=white, text=didarkgray
  },
  capbox/.style={
    font=\ttfamily\scriptsize, text=diblue, align=left
  },
  arr/.style={-{Stealth[length=6pt]}, very thick, color=diblue},
  narr/.style={
    font=\sffamily\scriptsize\bfseries, text=dired, midway, above
  },
]

\node[stage, fill=diblue!8] (dev) at (0,0) {
  \textbf{Developer}\\{\scriptsize (build time)}
};
\node[capbox, below=2pt of dev.south] {
  $S_{\text{declared}}$\\
  \{fs.read, fs.write,\\
  \phantom{\{}process.exec, \ldots\}
};

\node[stage, fill=diaccent!8] (op) at (3.8,0) {
  \textbf{Operator}\\{\scriptsize (provisioning)}
};
\node[capbox, below=2pt of op.south] {
  $\Smax \subseteq S_{\text{declared}}$\\
  \{fs.read,\\
  \phantom{\{}fs.write\}
};

\node[stage, fill=dipurple!8] (gar) at (7.6,0) {
  \textbf{GAR}\\{\scriptsize (issuance)}
};
\node[capbox, below=2pt of gar.south] {
  verifies $\subseteq$\\
  signs AIC with\\
  $\Smax$
};

\node[stage, fill=digreen!8] (rt) at (11.4,0) {
  \textbf{Runtime}\\{\scriptsize (per session)}
};
\node[capbox, below=2pt of rt.south] {
  $S_{\text{session}}$\\
  $= \Smax^A \!\cap\! \Smax^B$\\
  $\cap\; S_{\text{policy}}$
};

\draw[arr] (dev) -- (op) node[narr] {$\subseteq$};
\draw[arr] (op) -- (gar) node[narr] {$\subseteq$};
\draw[arr] (gar) -- (rt) node[narr] {$\cap$};

\node[font=\sffamily\small\itshape, text=dired, anchor=north]
  at (5.7, -2.5)
  {Each stage can only \textbf{narrow}, never widen.
   Authority flows left$\to$right via $\subseteq$ and $\cap$.};

\end{tikzpicture}

%% file: sections/accountability.tex
\section{Layer~3: Accountability \& Economics}
\label{sec:accountability}

Layers~0--2 establish who an agent is, what it can do, and whom it trusts.
Layer~3 closes the loop by ensuring that every agent action is
\emph{traceable}, \emph{attributable}, and \emph{economically accountable}.
This layer addresses threats T5 (payment and usage fraud) and T6 (action
repudiation) from \Cref{tab:threats}.

\subsection{Token-Usage Tracing}
\label{subsec:token-tracing}

LLM-powered agents consume computational resources (inference tokens, tool
invocations, storage) on behalf of their operators. In multi-agent
workflows, a single user request may trigger cascading agent interactions,
each consuming tokens from different LLM providers. Without identity-aware
metering, costs are attributed to API keys rather than to the agents (and
ultimately the humans) who authorized the work.

\sys binds token-usage records to cryptographic agent identity:

\begin{protocolbox}[Token-Usage Record]
\small
\begin{tabbing}
\hspace{2em}\= \hspace{10em}\= \kill
\> \textbf{agent\_id}       \> $\DIDag$ of the consuming agent \\
\> \textbf{session\_id}     \> session token ID from Layer~2 attestation \\
\> \textbf{provider}        \> LLM provider identifier \\
\> \textbf{model}           \> model name and version \\
\> \textbf{input\_tokens}   \> prompt token count \\
\> \textbf{output\_tokens}  \> completion token count \\
\> \textbf{timestamp}       \> UTC timestamp \\
\> \textbf{delegation\_chain} \> ordered list of AIC IDs from root to leaf \\
\> \textbf{signature}       \> $\Sign_{\Kpriv}(\text{record payload})$
\end{tabbing}
\end{protocolbox}

The delegation chain field enables \emph{cost attribution up the delegation
tree}: if agent~$C$ was delegated by agent~$B$, which was delegated by
agent~$A$ (the root), the usage record traces the cost back to~$A$'s
operator. Combined with the AIC's operator identity (OIDC-bound), this
enables precise per-user, per-agent cost allocation in multi-tenant
environments.

\subsection{Payment Primitives}
\label{subsec:payment}

As agents increasingly provide services to other agents (e.g., a code-review
agent charging per review, a data-analysis agent billing per query), the
Internet of Agents requires payment primitives that are identity-aware and
auditable.

\sys defines a minimal payment protocol framework built on Layer~0--2
primitives:

\begin{enumerate}[leftmargin=2em,itemsep=2pt]
  \item \textbf{Price advertisement:} An agent's registration record
    (Layer~1) includes optional pricing metadata (rate type, currency, unit).
  \item \textbf{Payment negotiation:} During mutual attestation (Layer~2),
    the requester and responder can negotiate payment terms as part of the
    session establishment. Payment terms are included in the session token
    metadata.
  \item \textbf{Service delivery with metering:} During the session, both
    token usage and service-specific metering events are recorded with
    identity-signed records (\S\ref{subsec:token-tracing}).
  \item \textbf{Settlement:} After service completion, the metering records
    serve as cryptographically verifiable invoices. Settlement can occur
    through traditional payment rails, escrow services, or---for environments
    that support it---on-chain settlement~\cite{agent-osi}.
\end{enumerate}

\sys is deliberately agnostic about the settlement mechanism: it provides the
\emph{identity and metering infrastructure} that any payment system requires,
without mandating a specific payment rail. This follows Principle~P3 (no
infrastructure lock-in). Notably, Google's Agent Payments Protocol
(AP2)~\cite{ap2-protocol} is a recent open specification for AI-driven
agent payments, focusing on the settlement rail and transaction flow. \sys and AP2 are complementary: AP2 defines \emph{how
agents pay}, while \sys defines \emph{who is paying whom and with what
authority}, binding every payment event to a verified cryptographic
identity and a capability-bounded session.

\subsection{Action Accountability}
\label{subsec:action-accountability}

Every action an agent takes---tool invocations, delegations, message
exchanges---generates an \emph{execution trace entry} that is
cryptographically signed by the agent's kernel-held private key:

\begin{protocolbox}[Trace Entry]
\small
\begin{tabbing}
\hspace{2em}\= \hspace{10em}\= \kill
\> \textbf{trace\_id}       \> unique identifier \\
\> \textbf{agent\_id}       \> $\DIDag$ of the acting agent \\
\> \textbf{action\_type}    \> tool\_call $|$ delegation $|$ message $|$ payment \\
\> \textbf{action\_params}  \> structured parameters of the action \\
\> \textbf{action\_result}  \> outcome or status \\
\> \textbf{session\_id}     \> session context (from Layer~2) \\
\> \textbf{timestamp}       \> UTC timestamp \\
\> \textbf{prev\_hash}      \> hash of the preceding trace entry (chain integrity) \\
\> \textbf{signature}       \> $\Sign_{\Kpriv}(\text{entry payload} \,\|\, \text{prev\_hash})$
\end{tabbing}
\end{protocolbox}

The \texttt{prev\_hash} field chains trace entries into a hash-linked log,
providing tamper evidence: any modification to
a historical entry breaks the hash chain from that point forward. The
signature binds each entry to its agent's cryptographic identity, providing
non-repudiation.

\parab{Key gains of kernel-mediated accountability}
What distinguishes \sys's accountability model from application-level
logging or blockchain-anchored ledgers is its reliance on \emph{trusted,
kernel-mediated cryptography}. The kernel (whether running locally or as a
remote service) provides a verifiable, isolated trust boundary, typically
anchored by hardware mechanisms like Trusted Execution
Environments (TEEs) and secure boot. Because the agent's private key $\Kpriv$ never leaves
this kernel, the architecture yields three key gains:
(1)~\emph{Decentralized non-repudiation}: agents can cryptographically prove
their execution history and attribute costs without relying on a global
consensus ledger or trusted third-party auditor.
(2)~\emph{Deployment ubiquity}: immutable audit trails can be maintained even
in air-gapped, edge, or resource-constrained environments where blockchain
nodes are economically or technically unviable.
(3)~\emph{Compromise containment}: because the application logic cannot access
the raw signing key or rewrite historical logs, an attacker compromising the
agent's LLM or memory cannot forge retroactive trace entries, guaranteeing
the integrity of the audit trail up to the exact moment of breach.

\parab{Contrast with \blocka}
In \blocka~\cite{blocka2a}, audit trails were anchored to a blockchain via
Merkle proofs, providing immutability backed by distributed consensus. \sys
achieves comparable tamper-evidence through identity-signed hash chains,
which work in any environment. For deployments requiring stronger
immutability guarantees, the hash chain root can be periodically anchored to
a public timestamping service or distributed ledger.

\subsection{Non-Repudiation}
\label{subsec:non-repudiation}

Non-repudiation in \sys rests on two pillars:

\begin{enumerate}[leftmargin=2em,itemsep=2pt]
  \item \textbf{Kernel-mediated signing:} The agent's private key
    $\Kpriv$ is held exclusively by the kernel. All signing operations
    (attestation challenges, session tokens, trace entries, usage records)
    are mediated by the kernel's signing API. The agent's application logic
    cannot forge signatures because it never has access to the raw key.
  \item \textbf{AIC chain binding:} Every signature can be verified against
    the agent's AIC, which chains back to the GAR root key. The verifier
    needs only the GAR's public key (a well-known trust anchor) to validate
    any trace entry from any agent in the ecosystem.
\end{enumerate}

Together, these mechanisms ensure that if a trace entry bears a valid
signature under an agent's $\Kpub$, and the agent's AIC chain verifies back
to the GAR root, then the named agent \emph{did} perform the recorded action.
The agent cannot repudiate it without claiming that the kernel was
compromised---a claim that can be evaluated against the key protection tier
(\Cref{tab:key-tiers}) and deployment attestation records.

%% file: sections/analysis.tex
\section{Security Analysis}
\label{sec:analysis}

A protocol suite spanning identity, discovery, trust negotiation, and accountability must be evaluated by whether its properties hold jointly under a defined attacker model. This section provides a rigorous, property-centric security evaluation of \sys. \Cref{tab:security-props} summarizes the security properties provided by \sys.

\subsection{Threat Model and Assumptions}
\label{subsec:attacker-model}

We assume a \emph{Dolev-Yao network adversary}~\cite{dolev-yao} capable of intercepting, replaying, and injecting messages across all network paths. Crucially, the adversary is augmented with \emph{LLM-level compromise capabilities}: it can control an agent's application logic, issue arbitrary API calls, and observe context data (e.g., via prompt injection or malicious memory retrieval). However, the attacker \emph{cannot} extract the agent's private key~$\Kpriv$ from the underlying kernel or compromise the kernel's signing API.

Our evaluation rests on three cryptographic and structural assumptions: (1)~\textbf{GAR Honesty}: The Global Agent Registry root key is uncompromised and accurately verifies developer/operator bindings during AIC issuance. (2)~\textbf{Kernel Integrity}: The agent's key-custody layer (the kernel) enforces delegation bounds and never signs payloads without authorization. (3)~\textbf{Cryptographic Soundness}: Ed25519 provides existential unforgeability (EUF-CMA) and $\Hash$ is collision-resistant. Graceful degradation assumptions, such as loose time synchronization and OIDC provider honesty, limit the scope of localized failures without causing systemic compromise.

\begin{table}[t]
\centering
\caption{Security properties of the \sys protocol suite. Threats refer to \Cref{tab:threats}.}
\label{tab:security-props}
\small
\setlength{\tabcolsep}{4pt}
\begin{tabularx}{\linewidth}{@{}
  >{\raggedright\arraybackslash}p{0.20\linewidth}
  >{\raggedright\arraybackslash}X
  >{\raggedright\arraybackslash}p{0.22\linewidth}
  >{\raggedright\arraybackslash}p{0.12\linewidth}@{}}
\toprule
\textbf{Property} & \textbf{Mechanism} & \textbf{Guarantee Level} & \textbf{Threats} \\
\midrule
Identity Integrity & Four-dimensional AIC binding, GAR root anchor & Cryptographic & T1, T4 \\
Capability Confinement & Monotonic attenuation chain, two-tier access control & Structural & T2 \\
Delegation Safety & Subset enforcement, depth bounds, tenant isolation & Crypto + Structural & T2, T3 \\
Discovery Integrity & Verifiable credentials, supply-chain validation & Verifiable Cred. & T1, T2 \\
Accountability & Kernel-mediated signing, hash-linked execution logs & Cryptographic & T5, T6 \\
\bottomrule
\end{tabularx}
\end{table}

\subsection{Property-Centric Security Evaluation}
\label{subsec:property-eval}

We prove that \sys satisfies its core security properties under the defined threat model, directly neutralizing threats T1--T6 (\Cref{tab:threats}).

\parab{Identity \& Authenticity (T1, T4)}
An adversary attempting to spoof an agent's identity (T1) must forge the Agent Identity Credential (AIC). The AIC requires four independently verified signatures: developer code-package digest, build pipeline attestation, OIDC operator binding, and the final GAR endorsement. Because each dimension is cryptographically bound and verified prior to GAR issuance, single-dimension compromise (e.g., a rogue OIDC provider) cannot yield full identity forgery. At runtime, mutual attestation defeats replay attacks by requiring fresh nonce signing, proving possession of $\Kpriv$. Furthermore, the GAR acts as a shared root of trust, enabling cross-domain verification (T4) via a single AIC chain, eliminating the need for pairwise trust agreements.

\parab{Capability Confinement \& Delegation Safety (T2, T3)}
\sys neutralizes capability escalation (T2) structurally. The monotonic attenuation chain ensures that permissions can only be narrowed at each delegation stage (developer $\to$ operator $\to$ GAR $\to$ session). For any chain $\AIC_0 \to \dots \to \AIC_n$, the invariant $\Smax^{\AIC_n} \subseteq \dots \subseteq \Smax^{\AIC_0}$ is enforced cryptographically; violating it requires forging a signature at a prior stage. At runtime, the two-tier access control model computes $S_{\text{session}} = S_{\text{policy}} \cap (\Smax^A \cap \Smax^B)$. Even if the application-level policy ($S_{\text{policy}}$) is compromised or misconfigured, it cannot grant access beyond the cryptographically verified infrastructure bound. Delegation abuse (T3) is prevented via depth bounding, strict validity capping, tenant isolation, and cascading revocation, ensuring delegation cannot be used as a capability-laundering channel.

\parab{Discovery Integrity (T1, T2)}
Existing discovery directories suffer from self-declared, unverified claims. \sys transforms discovery into a verify-then-interact boundary using Verifiable Credentials (VCs). An agent cannot advertise unauthorized skills because manifest VCs undergo strict verification: supply-chain signature checks, subject binding against the presenter's $\DIDag$, and permission alignment against the agent's $\Smax$. Consequently, discovery cannot serve as an implicit escalation path.

\parab{Accountability \& Non-repudiation (T5, T6)}
Action repudiation (T6) and usage fraud (T5) are mitigated without relying on distributed consensus. Every operational trace entry is signed by $\Kpriv$ under kernel mediation. Since application logic never possesses $\Kpriv$, an LLM-compromised agent cannot forge signatures. Furthermore, trace entries are linked via cryptographic hashes (\texttt{prev\_hash}); an attacker modifying past entries invalidates the chain. Hence, usage records remain cryptographically attributable up the delegation tree to a human-accountable operator.

\subsection{Security under Composition}
\label{subsec:composition}

A common pitfall in protocol design is composition failure, where security properties holding in isolation break when layers interact. \sys achieves composition safety through strict \emph{downward-only dependencies} and \emph{independent failure domains}. The guarantees of higher layers rely only on the invariants of lower layers, never the reverse. For instance, L2 session capability intersection relies on L0's AIC integrity; however, a misconfigured L2 application policy cannot widen L0's cryptographic capability boundary. Similarly, L3 accountability depends solely on L0's kernel-mediated key custody, remaining tamper-evident regardless of L1 discovery mechanisms or L2 trust negotiation flaws. This structural isolation ensures that a compromise at the application or discovery layer cannot weaken the cryptographic and identity invariants enforced by the infrastructure.

\subsection{Comparison and Residual Risks}
\label{subsec:comparison-risks}

\Cref{tab:scenario-comparison} illustrates \sys's structural guarantees compared to AgentMesh~\cite{agentmesh}, AIP~\cite{aip-protocol}, and \blocka~\cite{blocka2a} under specific failure scenarios. Where existing approaches rely on correct runtime policy configuration (e.g., AgentMesh) or omit protocol-level delegation attenuation (e.g., \blocka), \sys provides cryptographic bounds that fail closed independently of the policy engine.

\begin{table}[t]
\centering
\caption{Security guarantee comparison under failure scenarios. \textbf{$\checkmark$}~= guarantee holds, \textbf{$\times$}~= guarantee breaks, \textbf{$\triangle$}~= bounded break.}
\label{tab:scenario-comparison}
\small
\begin{tabularx}{\textwidth}{@{}
  >{\raggedright\arraybackslash}p{0.22\textwidth}
  *{4}{>{\raggedright\arraybackslash}X}@{}}
\toprule
\textbf{Scenario} & \textbf{\sys} & \textbf{AgentMesh}~\cite{agentmesh} & \textbf{AIP}~\cite{aip-protocol} & \textbf{\blocka}~\cite{blocka2a} \\
\midrule
PDP misconfigured to allow \texttt{tools:*} & $\checkmark$ (bound by $\Smax$ intersection) & $\times$ (single-PDP silently grants access) & $\triangle$ (depends on Datalog correctness) & $\checkmark$ (smart contract enforced) \\[6pt]
Operator credential compromised & $\triangle$ (developer + code signatures remain valid) & $\times$ (workload impersonation) & $\times$ (root issuer mints full capabilities) & $\triangle$ (no dev/operator separation) \\[6pt]
Child escalating beyond parent's $\Smax$ & $\checkmark$ (cryptographic subset enforcement) & $\triangle$ (runtime policy error enables escalation) & $\checkmark$ (attenuation at token level) & --- (no protocol-level delegation attenuation) \\[6pt]
Attacker rewrites past execution logs & $\checkmark$ ($\Kpriv$ kernel-mediated, hash chain) & $\triangle$ (application-level logging) & $\triangle$ (application-level logging) & $\checkmark$ (blockchain immutability) \\
\bottomrule
\end{tabularx}
\end{table}

\parab{Scope Boundaries}
Certain risks fall explicitly outside \sys's protocol scope. \sys secures \emph{identity and authorization limits}, not cognitive correctness. If an agent suffers a prompt injection that causes it to misuse its \emph{authorized} capabilities, this is an LLM-level cognitive failure rather than a protocol flaw. Furthermore, catastrophic infrastructure compromise (e.g., GAR root key theft) or hardware-level key extraction (bypassing the kernel) represent ultimate trust boundaries mitigated by external operational security measures, such as HSMs and Trusted Execution Environments (TEEs), rather than protocol-level invariants.

%% file: sections/related.tex
\section{Related Work}
\label{sec:related}

\parab{How we contrast \sys with prior work}
We organize the rapidly growing body of agent protocol and security
research into six clusters and contrast \sys with each using a uniform
three-step pattern: (i)~the goal or mechanism that \sys \emph{shares}
with the cluster, (ii)~the specific point of \emph{divergence}, named
in the canonical vocabulary of persistent identity,
capability-aware discovery, trust negotiation, and accountability, and
(iii)~the operationally observable \emph{consequence}---typically a
concrete attacker action or misconfiguration that \sys structurally
denies and the comparator catches only via runtime policy evaluation,
behavioral attestation, or operator vigilance. We use this pattern
because it forces every distinction claim to be grounded in mechanism
rather than rhetoric.

\parab{What \sys inherits}
\sys is not built from scratch. It inherits decentralized identifiers
and verifiable credentials from the W3C and SSI lineage, SPIFFE-style
workload attestation, OAuth/OIDC scoping vocabulary, capability tokens
in the UCAN/Biscuit lineage, the idea of monotonic attenuation as a
delegation discipline, signed software manifests, and tamper-evident
hash-chained audit. \sys's
contribution is the \emph{architecture that combines and structurally
enforces} these primitives within a single coherent four-layer trust
plane; we therefore foreground the conjunction rather than any
individual primitive.

\Cref{tab:comparison} compares \sys with representative approaches across
thirteen evaluation dimensions ordered to mirror the four-layer trust
substrate in \Cref{fig:protocol-stack}. \textit{Layer~0 (Agent
Identity)} uses rows~1--2 (\emph{persistent identity},
\emph{four-dimensional binding}). \textit{Layer~1 (Discovery)} uses
rows~3--4 (\emph{capability-aware discovery}, \emph{verifiable
manifests}). \textit{Layer~2 (Trust Negotiation)} uses rows~5--8 in the
same sequence as \S\ref{subsec:architecture}: \emph{mutual attestation},
\emph{monotonic attenuation}, \emph{two-tier access control}, then
\emph{delegation chains}. \textit{Layer~3 (Accountability)} uses
rows~9--12 in the same sequence as \S\ref{sec:accountability}:
\emph{token-usage tracing}, \emph{payment primitives}, \emph{action
accountability}, \emph{non-repudiation}. Row~13 is cross-cutting
\emph{deployment generality} (independent of any single layer).

\begin{table}[t]
\centering
\caption{Comparison of \sys with representative agent security approaches.
\textbf{$\bullet$}~= comprehensive support, \textbf{$\circ$}~= partial
support, \textbf{---}~= not addressed.
Rows follow the \sys four-layer suite (\Cref{fig:protocol-stack}): L0
Agent Identity~(1--2), L1 Discovery~(3--4), L2 Trust Negotiation~(5--8
in protocol flow order), L3 Accountability~(9--12), then deployment
generality~(13). The leftmost column labels the protocol layer for each
block; \emph{Cross} marks a cross-cutting deployment dimension.
For deployment generality:
\textbf{$\bullet$}~= cloud\,+\,edge\,+\,air-gapped,
\textbf{$\circ$}~= cloud-only or specific infrastructure,
\textbf{---}~= blockchain required.}
\label{tab:comparison}
{\normalsize
\setlength{\tabcolsep}{3pt}
\begin{tabular}{@{} >{\centering\arraybackslash}p{3.75em} l cccccccc@{}}
\toprule
\textbf{Layer} &
\textbf{Dimension} &
  \rotatebox{70}{\sys} &
  \rotatebox{70}{AgentMesh} &
  \rotatebox{70}{AIP} &
  \rotatebox{70}{HDP} &
  \rotatebox{70}{ANP} &
  \rotatebox{70}{Ag-OSI} &
  \rotatebox{70}{ZT-IAM} &
  \rotatebox{70}{\blocka} \\
\midrule
\multirow{2}{*}{L0} & Persistent identity          & $\bullet$ & $\bullet$ & $\circ$ & --- & $\circ$ & $\circ$ & $\bullet$ & $\bullet$ \\
                    & 4-dim binding                & $\bullet$ & $\circ$   & ---     & --- & ---     & ---     & $\circ$   & ---       \\
\midrule
\multirow{2}{*}{L1} & Capability-aware discovery   & $\bullet$ & $\circ$   & ---     & --- & $\circ$ & $\circ$ & $\circ$   & ---       \\
                    & Verifiable manifests         & $\bullet$ & ---       & ---     & --- & ---     & ---     & $\circ$   & ---       \\
\midrule
\multirow{4}{*}{L2} & Mutual attestation           & $\bullet$ & $\circ$   & $\circ$ & --- & $\circ$ & ---     & $\circ$   & $\circ$   \\
                    & Monotonic attenuation        & $\bullet$ & ---       & $\circ$ & --- & ---     & ---     & ---       & ---       \\
                    & Two-tier access control      & $\bullet$ & ---       & ---     & --- & ---     & ---     & ---       & $\circ$   \\
                    & Delegation chains            & $\bullet$ & $\circ$   & $\circ$ & $\bullet$ & --- & --- & $\circ$   & ---       \\
\midrule
\multirow{4}{*}{L3} & Token-usage tracing          & $\bullet$ & $\circ$   & ---     & --- & ---     & ---     & ---       & ---       \\
                    & Payment primitives           & $\bullet$ & ---       & ---     & --- & ---     & $\bullet$ & ---     & ---       \\
                    & Action accountability        & $\bullet$ & $\bullet$ & $\circ$ & $\circ$ & --- & $\circ$ & $\circ$ & $\bullet$ \\
                    & Non-repudiation              & $\bullet$ & $\circ$   & $\circ$ & $\circ$ & --- & ---     & ---       & $\bullet$ \\
\midrule
\emph{Cross}        & Deployment generality        & $\bullet$ & $\circ$   & $\bullet$ & $\bullet$ & $\circ$   & ---   & $\circ$   & ---       \\
\bottomrule
\end{tabular}%
}
\end{table}

\subsection{Communication Stacks vs.\ Security-First Substrates}

A first cluster of work designs communication-first stacks for the
Internet of Agents. Fleming et al.~\cite{ioa-layered} propose reference layers for agent
communication (L8) and semantics (L9) atop TCP/IP (outside the classical OSI
model);
Agent-OSI~\cite{agent-osi} proposes a six-layer decentralized stack with
identity, settlement, and provenance; ACPS~\cite{acps} defines
registration, discovery, interaction, and tooling protocols; Coral
Protocol~\cite{coral-protocol} provides open infrastructure for
communication, coordination, trust, and payments; the OpenAgents
Network Model~\cite{openagents-network-model} defines an event-centered
network abstraction with scoped networks, routable addresses, mods,
resources, discovery, and transport bindings; Du
et al.~\cite{ai-agent-comm} analyze agent communication from five
classical Internet-architecture perspectives; the IoA framework of Chen
et al.~\cite{ioa-weaving} and the survey by Wang
et al.~\cite{ioa-fundamentals} provide ecosystem perspectives; and
Google's Agent Payments Protocol (AP2)~\cite{ap2-protocol} is a prominent
open effort focused on AI-driven agent payments. A parallel cluster of comparative
surveys~\cite{agent-protocol-survey,agent-protocol-survey-ehtesham,
agentic-ai-frameworks,mcp-x-a2a,agentic-web-mdpi,a2h,agent-discovery-ioa}
catalog and compare these efforts (including landscape surveys with limited
security depth, such as~\cite{agentic-web-mdpi,a2h}); across these surveys, security is
typically treated as one dimension among several rather than as a dedicated,
cryptographically bound trust layer.

\parab{Relation to \sys}
\sys shares with these works the goal of supporting interoperable
multi-agent collaboration; it diverges by being a security-first
protocol suite that is agnostic to the underlying communication stack,
and is therefore designed to \emph{complement} rather than compete with
any of the architectures above. In particular, OpenAgents is best
understood as a network model rather than an in-depth security or
verifiability design: its verification levels and guard mods name where
authentication, access control, and rate limiting can occur, but they do
not specify AIC-style multi-dimensional identity binding, signed
capability manifests, monotonic attenuation, two-tier authorization, or
tamper-evident audit. Similarly, while ANP~\cite{anp-whitepaper} uses
W3C DIDs for discovery and mutual verification, its published design does
not specify structured delegation chains or a tamper-evident accountability
layer comparable to \sys; attestation and access control are lighter than
cryptographically bound, challenge-response attestation with monotonic
capability attenuation. Agent-OSI~\cite{agent-osi} provides comprehensive
layering and prototypes on-chain escrow and verification in its reference
implementation, limiting deployment generality where blockchain is required. The observable consequence is that \sys can layer
on top of MCP, A2A, ANP, ACPS, Agent-OSI, or OpenAgents-style event
networks without changes to the host protocol's wire format.

\subsection{Single-Dimension IAM vs.\ Multi-Dimensional Binding}
\label{subsec:rel-iam}

A second cluster retrofits human-IAM primitives onto agent flows. The
OpenID Foundation's strategic agenda~\cite{iam-agentic-ai} catalogs the
gaps in extending OAuth~2.0 and OpenID Connect to agentic
authorization; OIDC-A~\cite{oidc-a} is the most concrete OIDC extension
to date, defining standard claims for agent identity, attestation, and
delegation chains. South et
al.~\cite{authenticated-delegation,authenticated-delegation-icml}
extend OAuth~2.0 with agent-specific credentials and natural-language
permission scoping. Bhushan et al.~\cite{auth-taxonomy} provide a
five-pattern taxonomy organizing SPIFFE/SPIRE, OAuth~2.0, OIDC, Token
Exchange, DPoP, CIBA, and decentralized identity by interaction type
(user-to-agent, orchestrator-to-agent, agent-to-internal,
agent-to-external, cross-domain federation). On the
decentralized-identity side, Aydeger and
Zeydan~\cite{decentralized-llm-identity} integrate SSI with LLM agents
on a blockchain backend; AGNTCY Identity~\cite{agntcy-identity} and
BillionsNetwork~\cite{billions-agent-identity} provide open-source
toolkits using W3C VCs and the iden3 protocol respectively.

\parab{Relation to \sys}
\sys shares with this cluster the use of standardized credential
formats and the goal of cross-domain agent authentication. It diverges
by defining an agent-native AIC lifecycle whose four-dimensional
binding (developer, code package, operator, context) is signed at
issuance, rather than treating the agent as a single OAuth client or a
single VC holder. The consequence is that an attacker who steals an
operator's OAuth client secret or who phishes a holder key cannot
silently substitute a different code package or claim a different
developer, because each dimension carries its own independently
verifiable signature---a property no OAuth or single-holder DID
extension provides.

\subsection{Directory Discovery vs.\ Capability-Aware Manifests}

A third cluster builds zero-trust identity and discovery infrastructure
for agents. The DID/VC + Agent Name Service strand---Huang
et al.'s zero-trust framework with DIDs, VCs, ANS, and ZKPs~\cite{zt-agentic-ai},
the ANS itself as a DNS-inspired PKI directory~\cite{ans}, and Garzon
et al.'s ledger-anchored DID/VC prototype for cross-domain LLM-agent
authentication~\cite{did-vc-agents}---focuses on \emph{discovery and
authentication}. The SPIFFE-on-Kubernetes strand---Huang and Hughes's
Springer chapter on agentic AI identity
security~\cite{agentic-ai-identity-security}, Pappu et al.'s
SPIFFE-based zero-trust authentication~\cite{spiffe-agents},
Bhushan's explainable zero-trust framework~\cite{zt-explainable}, and
Palavali's SSI-for-microservices framework~\cite{ssi-microservices}---focuses
on \emph{workload attestation and short-lived credentials}. A third
strand frames the problem at the enterprise governance layer:
Ramachandran and Mishra's identity-aware
governance~\cite{identity-aware-governance} explicitly proposes
ambient-authority elimination, infrastructure-level policy
enforcement, sequence-aware authorization, independent action
verification, and hallucination-aware audit, and grounds these in
documented production incidents.

\parab{Relation to \sys}
\sys shares with this cluster the zero-trust posture and the use of
DID/SPIFFE-style cryptographic identity. It diverges in two ways. First,
the DID/VC works typically bind only the \emph{holder} dimension and
the SPIFFE works only the \emph{workload-instance} dimension; \sys's
AIC binds developer, code package, operator, and context as
independently signed dimensions, so the four-dimensional binding
strictly subsumes both. Second, the enterprise-governance proposals
share \sys's two-tier intuition (separating infrastructure from
application policy) but realize it through a single PDP at the
infrastructure layer; \sys realizes it as two independent evaluation
paths with independent failure semantics. The consequence is that a
PDP misconfiguration in the application tier cannot weaken
cryptographic AIC verification in the infrastructure tier, and vice
versa---each tier fails closed independently.

\subsection{Policy-Evaluated vs.\ Structurally-Attenuated Delegation}
\label{subsec:rel-delegation}

A fourth cluster focuses specifically on \emph{delegation chains} as
first-class protocol artifacts. We discuss this cluster in detail
because it is where the monotonic-capability-attenuation primitive of
\sys has the closest competitors.

AIP~\cite{aip-protocol} introduces Invocation-Bound Capability Tokens
(IBCTs) that fuse identity, attenuated authorization, and provenance
into an append-only token chain, with compact (signed JWT) and chained
(Biscuit/Datalog) wire formats and transport bindings for MCP, A2A,
and HTTP. LDP~\cite{ldp-protocol} extends this with rich delegate
identity cards carrying quality hints, governed sessions, and trust
domains as protocol-level boundaries. HDP~\cite{hdp} (an IETF
Internet-Draft) provides a lightweight Ed25519 hop-chain dedicated to
\emph{human delegation provenance} in agentic systems. Saavedra's
framework~\cite{saavedra-delegation} introduces Delegation Grants
(DGs)---first-class authorization artifacts with explicitly enforced
scope reduction---together with a Canonical Verification Context, a
Trust Gateway, and optional blockchain anchoring. South et
al.'s authenticated delegation~\cite{authenticated-delegation,
authenticated-delegation-icml} extends OAuth~2.0/OIDC with
agent-specific credentials and natural-language scoping for delegation
chains.

\parab{Relation to \sys}
All five works share with \sys the goal of bounding what a delegated
agent may do and producing an auditable chain of authority; AIP and
Saavedra's DGs go furthest by making attenuation a structural property
of the credential rather than only a runtime check. The divergences
are mechanism-specific. AIP's attenuation is \emph{invocation-scoped}
and \emph{policy-evaluated}: each IBCT carries Datalog rules
re-evaluated at every hop, and the verifier must run a Datalog engine
to reject a malformed token. \sys's attenuation is
\emph{lifecycle-scoped} and \emph{intersection-evaluated}: the
capability boundary is signed into the AIC at issuance and every
subsequent narrowing reduces to set-intersection plus signature
verification---no Datalog runtime, no per-hop policy interpretation.
Neither AIP nor Saavedra's DGs separate \emph{developer} from
\emph{operator} as distinct issuance stages, so a compromised AIP
root issuer or a compromised DG-issuing trust gateway can mint a
credential with the full upstream capability set; \sys's
four-dimensional binding makes the developer-declared boundary and
the operator-provisioned subset two distinct cryptographic stages, so
an operator-tier compromise structurally cannot escalate beyond the
developer-tier boundary. HDP and the OAuth-extension delegation
proposals address human-to-agent provenance and OAuth compatibility
respectively, but do not extend monotonic attenuation across the
multi-stage agent supply chain. The observable consequence is that
the two attack categories that AIP's chained model uniquely catches
(delegation-depth violations and audit evasion via empty
context~\cite{aip-protocol}) are also caught by \sys, plus a third
category---operator-tier capability widening---that AIP's
single-issuer model does not catch.

\subsection{Conformance Testing vs.\ Cryptographic Enforcement}

A fifth cluster articulates security principles and threat models for
the agentic ecosystem rather than full protocols. AgentRFC~\cite{agentrfc}
is the closest principles framework to \sys: it defines a six-layer
Agent Protocol Stack analogous to ITU-T~X.800 for OSI, formalizes
eleven security principles as TLA+ invariants with an explicit
property taxonomy (spec-mandated, spec-recommended, AASM-hardening,
APS-completeness), and introduces the Composition Safety
principle---the observation that security properties holding for
individual protocols can break when protocols are composed through
shared infrastructure. The AgentConform tooling extracts normative
clauses into a typed Protocol IR and model-checks the resulting TLA+
model. Anbiaee et al.~\cite{anbiaee-threat-model} present
comparative threat modeling across MCP, A2A, Agora, and ANP and
enumerate twelve protocol-level risks. Wibowo and
Polyzos~\cite{safe-ioa} argue that agentic safety is an architectural
principle rather than an add-on, and bottom-up deconstruct single-,
multi-, and interoperable-multi-agent stacks. Chaffer's ``Know Your
Agent''~\cite{know-your-agent} proposes a governance frame centered on
identity verification, behavioral monitoring, and accountability; the agent discovery survey by Guo
et al.~\cite{agent-discovery-ioa} introduces a two-stage
capability-discovery framework with semantic modeling.

\parab{Relation to \sys}
\sys shares with AgentRFC the diagnosis that agent protocols need a
principled security framework, and we adopt the Composition Safety
principle in our threat model (\S\ref{subsec:threat-model}). The
divergence is that AgentRFC's invariants are \emph{model-checked}
properties intended for conformance testing of independent protocols,
whereas \sys's invariants are \emph{cryptographically structural}---the
monotonic capability attenuation chain enforces the bound at the
credential level, not at the spec-conformance level. The consequence
is complementary: AgentRFC can flag where existing specs leave
Composition Safety gaps; \sys provides a credential format that
closes those gaps by construction. The threat-modeling and
governance-framework works are likewise complementary---they identify
risks that \sys is designed to neutralize architecturally rather than
articulate.

\subsection{Single-Plane Overlays vs.\ Two-Tier Authorization}
\label{subsec:rel-overlay}

The sixth cluster contains the two industry- or industry-adjacent
efforts that share \sys's framing as a trust layer added on top of
existing communication protocols: Microsoft's
AgentMesh~\cite{agentmesh} and Huang et al.'s unified zero-trust
architecture for the agentic web~\cite{zt-unified}.

\parab{AgentMesh}
AgentMesh, released as part of the Agent Governance Toolkit, is marketed in
its documentation as ``SSL for AI agents'' (Public Preview). It organizes its architecture into
four layers: (i)~an identity and zero-trust core using Agent CA with
SPIFFE/SVID identities and Ed25519 or ML-DSA-65 signatures; (ii)~a
trust and protocol bridge that translates between A2A, MCP, and IATP
with capability scoping; (iii)~a compliance and policy plane; and
(iv)~a reward and learning engine. Project documentation claims alignment
with the OWASP Agentic Top~10 categories and ships an MCP proxy.

\parab{Relation to \sys}
\sys and AgentMesh share the high-level goal of adding a trust layer
to the multi-agent ecosystem and both use signed credentials for
identity. They differ in five mechanism-level ways, each with a
concrete observable consequence:
\begin{enumerate}[leftmargin=2em,itemsep=2pt]
  \item \textbf{Identity model.} AgentMesh's SPIFFE-based identity
    binds a key to a single workload instance; \sys's AIC binds
    developer, code package, operator, and operational context as
    four independently signed dimensions. \emph{Consequence:} an
    attacker who compromises an AgentMesh workload signs valid SVIDs
    for the workload's identity but cannot prove anything about its
    code provenance; under \sys, the same compromise cannot
    impersonate a different developer or substitute a different code
    package because the developer and code-package signatures are
    independent of the operator-tier credential.
  \item \textbf{Bound enforcement: structural vs.\ policy-based.}
    AgentMesh enforces least-privilege through a runtime policy
    engine: a misconfigured policy rule (e.g., a tenant operator
    accidentally granting \texttt{tools:*}) can silently grant
    excessive access. \sys's monotonic capability attenuation chain
    encodes the bound at credential issuance, so the same policy edit
    cannot widen capability beyond the developer-declared boundary
    regardless of any subsequent runtime policy decision.
    \emph{Consequence:} \sys remains safe under PDP misconfiguration
    in a way AgentMesh does not.
  \item \textbf{Access-control architecture.} AgentMesh uses a single
    policy plane that conflates cryptographic verification with
    declarative authorization. \sys's two-tier access control keeps
    infrastructure-tier checks (AIC validity, capability
    intersection) on a separate evaluation path from application-tier
    declarative rules. \emph{Consequence:} an application-tier
    regression cannot weaken cryptographic checks, and an
    infrastructure-tier credential lapse cannot bypass per-agent
    authorization---each failure mode is independently auditable.
  \item \textbf{Protocol relationship.} AgentMesh ships explicit
    A2A/MCP/IATP protocol translators; \sys is protocol-agnostic and
    provides trust primitives that any host protocol can bind to.
    \emph{Consequence:} \sys does not require a translator update to
    support a new host protocol, and protocol additions do not
    expand the trust-layer attack surface.
  \item \textbf{Scope.} AgentMesh embeds a reward and learning engine
    for adaptive governance. \sys deliberately confines itself to
    the trust and accountability layer and treats adaptive behavior
    as orthogonal. \emph{Consequence:} \sys's verifier semantics
    remain pure cryptographic checks, simplifying formal verification
    and audit.
\end{enumerate}

\parab{Unified zero-trust architecture (Huang et al.)}
\sys's closest \emph{academic} comparator is Huang
et al.'s unified zero-trust architecture~\cite{zt-unified}, which
likewise proposes a unified zero-trust architecture for the agentic web.
It builds on DIDs, VCs, and ANS for identity and discovery and adds
three mechanisms: Trust-Adaptive Runtime Environments (TARE), Causal
Chain Auditing, and Dynamic Identity with Behavioral Attestation. It
shares with \sys the diagnosis that the agentic web needs a unified
trust layer; it diverges by enforcing trust through \emph{runtime
behavioral attestation and trust scoring} rather than through
issuance-time cryptographic attenuation. The observable consequence is
that an agent whose behavior has not yet diverged from baseline
receives a high trust score from TARE even if its capability
boundary has been silently widened upstream; under \sys the same
agent's AIC-bounded capability set is unchanged regardless of
behavioral history.

\subsection{Our Prior Work: \blocka}

\blocka~\cite{blocka2a} introduced a unified multi-agent trust
framework combining DIDs for cross-domain authentication,
blockchain-anchored ledgers for immutable audit, and smart contracts
for dynamic access control. It demonstrated effective defense against
diverse MAS attacks and practical integration with A2A~\cite{a2a-protocol}.

\parab{Evolution to \sys}
\sys preserves \blocka's core security ambitions (unified trust,
immutable audit, dynamic access control) while making three key
advances: (1)~removing the blockchain dependency in favor of
general-purpose cryptographic primitives, broadening deployment to
edge and air-gapped environments; (2)~introducing layer-aligned
primitives for persistent identity, capability-aware discovery, trust
negotiation, and accountability, whereas \blocka{} approximated
identity and authorization only at the smart-contract policy level; and
(3)~adding DID-bound skill/tool manifest VCs, payment, and token-usage
tracing capabilities that \blocka{} did not address. \sys can be
viewed as the generalization of \blocka{} from a blockchain-specific
realization to a deployment-agnostic protocol suite.

\subsection{Summary: the Joint-Coverage Argument}
\label{subsec:rel-joint-coverage}

Each of \sys's primitives is anticipated by at least one prior work
surveyed above: DID/VC and SPIFFE precede the AIC's identity
machinery; UCAN, Biscuit, AIP IBCTs, and Saavedra DGs precede the
attenuation discipline; AgentMesh's policy plane and the enterprise
governance frameworks precede the two-tier intuition; AgentRFC
precedes the principled-invariants framing; and \blocka{} precedes
the unified-trust ambition. \sys's contribution is twofold.

First, \sys \emph{deepens} each primitive beyond its nearest
predecessor. Where SPIFFE, OAuth, and DIDs each bind a single
identity dimension, the AIC binds four independently signed
dimensions so that compromising one tier cannot impersonate another.
Where A2A Agent Cards and ANP's DID-based discovery treat skill and tool
advertisements as self-declared descriptions, \sys represents each
skill and tool as a DID-bound Verifiable Credential that is verified
for supply-chain authenticity, subject binding, and permission
alignment at discovery time---making discovery a trust-establishing
security boundary rather than a directory lookup.
Where AIP and Saavedra DGs attenuate capabilities via per-hop
Datalog or runtime policy, \sys's monotonic chain signs the
capability boundary at issuance and reduces verification to signature
checking, so the bound holds even when the policy engine is wrong.
Where AgentMesh and the enterprise governance frameworks route
cryptographic checks and application policy through a single PDP,
\sys's two-tier split keeps them on independent evaluation paths
with independent failure semantics, so a regression in one tier
cannot weaken the other. And where \blocka{} anchors audit trails
to blockchain consensus, \sys achieves tamper-evident
non-repudiation via kernel-mediated signing that works in air-gapped
and resource-constrained environments without a distributed ledger.

Second, \sys provides the \emph{conjunction}: no prior single
architecture jointly enforces persistent identity with four-dimensional
binding, capability-aware discovery through DID-bound Verifiable
Credential manifests, trust negotiation that combines lifecycle-scoped
monotonic capability attenuation with two-tier access control, and
kernel-mediated cryptographic audit trails.
\Cref{tab:comparison} makes both contributions visible across
thirteen evaluation dimensions: every prior approach achieves at
most a strict subset of \sys's coverage in breadth, and the
per-primitive depth gap is most pronounced precisely where
structural guarantees matter---under operator compromise, PDP
misconfiguration, or cross-protocol composition.

%% file: sections/discussion.tex
\section{Discussion \& Future Directions}
\label{sec:discussion}

In this section, we discuss the limitations of the \sys framework and outline key directions for future research. We first acknowledge the boundaries of this positioning paper and the practical challenges of deploying a global trust anchor. We then highlight promising avenues for future work, ranging from formal verification and privacy-preserving extensions to scalable infrastructure and integration with existing agent frameworks.

\subsection{Limitations}

\parab{Positioning scope}
\label{subsec:scope}
\sys is a positioning paper that establishes the conceptual framework,
protocol architecture, and design rationale. A complete protocol
specification would additionally require formal proofs of security
properties, quantitative performance evidence, implementation detail, and
adversarial validation; we defer these to companion work. Concretely, the
following aspects are explicitly out of scope here:

\begin{itemize}[leftmargin=2em,itemsep=2pt]
  \item \textit{Formal verification:} TLA+ or ProVerif models of the
    attestation and delegation protocols.
  \item \textit{Performance evaluation:} Latency and throughput benchmarks
    under realistic multi-agent workloads.
  \item \textit{Implementation details:} A companion \deepkernel system
    paper will describe the full implementation including kernel
    architecture, eBPF enforcement, and perception/cognition layers.
  \item \textit{Adversarial evaluation:} Red-team exercises and attack
    simulation against the protocol suite.
\end{itemize}

\parab{GAR as a trust anchor}
The Global Agent Registry serves as a centralized (or federated) trust
anchor, analogous to a Certificate Authority. This introduces the same
trade-offs as traditional PKI: the GAR must be highly available, its
compromise would be catastrophic, and cross-GAR trust requires federation
agreements. Decentralized alternatives (e.g., blockchain-backed GARs, web
of trust models) can mitigate these risks but introduce latency and
governance complexity.

\parab{Adoption bootstrapping}
The value of \sys increases with adoption (network effects). Early
deployments may face a chicken-and-egg problem where few agents support
\sys's identity layer, reducing the incentive for others to adopt. We
envision a phased adoption path: Layer~0 (identity) can be adopted
independently of Layers~1--3, providing immediate value for agent
authentication even when only a subset of the ecosystem participates.

\parab{Semantic tag governance}
Layer~1's semantic tagging relies on an open vocabulary. Without
governance, semantic tags may become fragmented or misleading. We
anticipate that domain-specific tag ontologies will emerge organically
(similar to Docker image tags or npm package keywords) and can be curated
by registry operators.

\subsection{Future Directions}

\parab{Formal verification}
The attestation and delegation protocols lend themselves to formal
verification using tools such as TLA+~\cite{agentrfc} for safety
properties and ProVerif or Tamarin for cryptographic protocol analysis.
Formalizing the monotonic attenuation chain as a lattice-theoretic
invariant would strengthen the structural guarantee claims.

\parab{Attestation deployment topologies}
The mutual attestation protocol (\S\ref{subsec:attestation}) is
parameterized by the relationship between \KerA and \KerB, yielding three
deployment modes with distinct trust and performance profiles.
\emph{Co-located attestation} (\KerA${} = {}$\KerB) is the lowest-latency
option, natural for \deepkernel-managed multi-agent orchestration on a
single host; the entire challenge-response collapses to kernel-internal
operations with no network round-trip.
\emph{Remote-mediated attestation} interposes a Trusted Attestation
Service (\TAS)---architecturally identical to a local agent kernel but
deployed as a cloud endpoint---that coordinates nonce exchange, caches GAR
verification results, and issues session tokens. The \TAS plays a role
analogous to an OIDC Provider in human web SSO: it centralizes session
establishment at the cost of introducing a single point of trust that must
itself be highly available and key-protected.
\emph{Direct peer-to-peer attestation} eliminates the intermediary: each
kernel signs its own nonce and independently verifies the counterpart's
signature via the GAR. This mode is fully decentralized but incurs at
least two independent GAR round-trips per attestation; AIC caching with
revocation-push notifications can amortize the cost.
The three modes can coexist within a single ecosystem---co-located for
intra-device workflows, mediated for organizational clusters, and direct
P2P for open cross-organization interactions---and the choice is
determined by the deployment context rather than by the protocol itself.

\parab{Privacy-preserving extensions}
The current design exposes the requester's full AIC attributes to the
responder during Layer~2 attestation. Zero-knowledge proofs (ZKPs) could
enable selective disclosure: an agent could prove it possesses a required
capability or meets an IAL threshold without revealing its full identity.
This is particularly relevant for privacy-sensitive domains such as
healthcare and finance.

\parab{Integration with agent frameworks}
\sys is designed to be framework-agnostic, but practical adoption requires
SDK integrations with popular agent frameworks (LangGraph, AutoGen,
CrewAI, Semantic Kernel) and vendor SDKs (OpenAI Agents
SDK~\cite{openai-agents-sdk}, Google ADK~\cite{google-adk}). We envision
thin adapter libraries that bridge each framework's identity and tool
invocation APIs to \sys's protocol primitives. AgentMesh's approach of
building explicit protocol translators for A2A, MCP, and IATP (in the same
Agent Governance Toolkit monorepo as AgentMesh)~\cite{agentmesh} suggests a
pragmatic integration pattern that \sys could adopt for its initial deployment
targets.

\parab{Scalability of the GAR}
As the agent population grows to millions or billions, the GAR must scale
correspondingly. Horizontal sharding by tenant or geographic region,
caching at the kernel level (with revocation push notifications), and
eventual consistency models for non-critical metadata are all viable
strategies that need empirical evaluation.

\parab{Cross-GAR federation}
In a multi-stakeholder ecosystem, multiple GARs will coexist (enterprise
GARs, cloud-provider GARs, open-community GARs). Cross-GAR trust
establishment requires mutual recognition protocols analogous to
cross-certification in PKI or trust federation in SAML. Designing these
protocols is a natural extension of \sys's architecture.

\parab{Agentic payment infrastructure}
Layer~3's payment primitives are deliberately minimal. A full agentic
payment infrastructure would need to address escrow for long-running
tasks, dispute resolution, quality-of-service guarantees, and
micro-payment efficiency. Google's AP2~\cite{ap2-protocol} is a leading open effort in this
direction, focusing on the payment rail and transaction flow. We view \sys's identity and metering layer as the
foundation upon which AP2-style settlement protocols can be built,
providing the ``who is paying whom and with what authority'' semantics that
payment rails alone cannot supply.

\parab{Post-quantum readiness}
AgentMesh~\cite{agentmesh} already lists ML-DSA-65 alongside Ed25519 as a
supported signature scheme. As NIST post-quantum standards mature, \sys's
AIC and attestation protocols should be extended to support PQ signature
algorithms (e.g., ML-DSA, SLH-DSA) as drop-in replacements for Ed25519,
preserving the structural security properties while future-proofing the
cryptographic substrate.

%% file: sections/conclusion.tex
\section{Conclusion}
\label{sec:conclusion}

The Internet of Agents will not become secure merely by standardizing how
agents exchange messages. The harder problem is whether independently built
agents can make interoperable trust decisions: is this agent the entity it
claims to be, are its advertised skills and tools genuine, can its authority
only decrease as it is delegated, and can its actions later be attributed
without ambiguity? This paper has argued that these questions require a
trust substrate, not another communication protocol.

\sys provides such a substrate as a four-layer protocol suite spanning
Persistent Identity, Discovery, Trust Negotiation, and Accountability. Its
principal contribution is the conjunction of four layer-aligned primitives:
Agent Identity Cards with four-dimensional binding; capability-aware
discovery through DID-bound skill/tool manifest VCs; trust negotiation that
combines monotonic capability attenuation with two-tier access control; and
kernel-mediated cryptographic audit trails. Each primitive has antecedents
in existing identity, delegation, policy, or audit systems. What is missing
from prior work, and what \sys supplies, is their joint enforcement across
the full agent lifecycle: from issuance and discovery, through negotiation
and delegation, to metering and non-repudiation.

This architecture deliberately remains communication-protocol agnostic.
Rather than competing with MCP, A2A, ANP, AG-UI, or future agent protocols,
\sys defines the trust objects those protocols can carry: AIC capability
boundaries, manifest VCs, session tokens, and signed traces. This separation
is what makes the design deployable across cloud, edge, and air-gapped
settings without relying on a single runtime, policy engine, registry
topology, or distributed ledger.

As a positioning paper, \sys establishes the conceptual framework and
protocol architecture; formal verification, performance benchmarks, and a
companion \deepkernel systems paper are left to ongoing work. The claim of
this paper is therefore precise: communication interoperability is necessary
for agents to talk, but secure interoperability is necessary for agents to
act across organizational boundaries. \sys is a step toward making the
underlying trust layer explicit, composable, and verifiable.

%% file: bibliography/refs.bib
@article{ioa-layered,
  title   = {A Layered Protocol Architecture for the Internet of Agents},
  author  = {Fleming, Charles and Muscariello, Luca and Pandey, Vivek and others},
  journal = {arXiv preprint arXiv:2511.19699},
  year    = {2025},
}

@article{ioa-fundamentals,
  title   = {Internet of Agents: Fundamentals, Applications, and Challenges},
  author  = {Wang, Yuntao and Guo, Song and Pan, Yutong and Su, Zhou and Chen, Fangmin and others},
  journal = {IEEE Transactions on Cognitive Communications and Networking},
  year    = {2025},
  note    = {arXiv:2505.07176; accepted by IEEE TCCN},
}

@article{agent-osi,
  title   = {{Agent-OSI}: A Layered Protocol Stack Toward a Decentralized Internet of Agents},
  author  = {Xu, Wenxin and Wang, Taotao and Xia, Yihan and Zhang, Shengli and Liew, Soung Chang},
  journal = {arXiv preprint arXiv:2602.13795},
  year    = {2026},
}

@article{ioa-weaving,
  title   = {Internet of Agents: Weaving a Web of Heterogeneous Agents for Collaborative Intelligence},
  author  = {Chen, Weize and You, Ziming and Li, Ran and Guan, Yitong and Qian, Chen and Zhao, Chenyang and others},
  journal = {arXiv preprint arXiv:2407.07061},
  year    = {2024},
}

@article{coral-protocol,
  title   = {Coral Protocol: Open Infrastructure Connecting the Internet of Agents},
  author  = {Georgio, RJ and Forder, C and Deb, S and Rahimov, A and others},
  journal = {arXiv preprint arXiv:2505.00749},
  year    = {2025},
}

@article{acps,
  title   = {{ACPS}: Agent Collaboration Protocols for the Internet of Agents},
  author  = {Li, Chao and Wu, Jiaxing and Du, Qiong and Yu, Shui and Zou, Rui and Yu, Kan and others},
  journal = {arXiv preprint arXiv:2505.13523},
  year    = {2025},
}

@misc{openagents-network-model,
  title        = {{OpenAgents} Network Model},
  author       = {{OpenAgents}},
  year         = {2026},
  howpublished = {\url{https://openagents.org/docs/zh/concepts/openagents-network-model}},
  note         = {Version 1.0; updated May 8, 2026},
}

@article{ai-agent-comm,
  title   = {{AI} Agent Communication from Internet Architecture Perspective: Challenges and Opportunities},
  author  = {Du, Chenguang and Wang, Chao and Chao, Yihan and Xie, Xin and Cui, Yong},
  journal = {arXiv preprint arXiv:2509.02317},
  year    = {2025},
}

@article{agent-protocol-survey,
  title   = {A Survey of {AI} Agent Protocols},
  author  = {Yang, Yingxuan and Chai, Huayi and Song, Yuqi and Qi, Shuai and Wen, Muning and Li, Na and others},
  journal = {arXiv preprint arXiv:2504.16736},
  year    = {2025},
}

@article{agent-protocol-survey-ehtesham,
  title   = {A Survey of Agent Interoperability Protocols: {MCP}, {ACP}, {A2A}, and {ANP}},
  author  = {Ehtesham, Abul and Singh, Aditi and Gupta, Gaurav Kumar and Kumar, Sandeep},
  journal = {arXiv preprint arXiv:2505.02279},
  year    = {2025},
}

@article{agentic-ai-frameworks,
  title   = {Agentic {AI} Frameworks: Architectures, Protocols, and Design Challenges},
  author  = {Derouiche, Hana and Brahmi, Zaki and Mazeni, Haithem},
  journal = {arXiv preprint arXiv:2508.10146},
  year    = {2025},
}

@article{mcp-x-a2a,
  title   = {A Study on the {MCP} x {A2A} Framework for Enhancing Interoperability of {LLM}-Based Autonomous Agents},
  author  = {Jeong, Cheonsu},
  journal = {arXiv preprint arXiv:2506.01804},
  year    = {2025},
}

@article{agentic-web-mdpi,
  title   = {{AI} Agent Communications in the Future Internet---Paving a Path Toward the Agentic Web},
  author  = {Duan, Qiang and Lu, Zhihui},
  journal = {Future Internet (MDPI)},
  volume  = {18},
  number  = {3},
  pages   = {171},
  year    = {2026},
}

@article{a2h,
  title   = {{A2H}: Agent-to-Human Protocol for {AI} Agent},
  author  = {Liang, Zhuo and Cui, Erzhuo and Wei, Qian and She, Rui and Li, Ting and Guo, Minghui and others},
  journal = {arXiv preprint arXiv:2602.15831},
  year    = {2026},
}

@article{agent-discovery-ioa,
  title   = {Agent Discovery in Internet of Agents: Challenges and Solutions},
  author  = {Guo, Song and Wang, Yuntao and Su, Zhou and Pan, Yutong and Hu, Qimei and Luan, Tom H.},
  journal = {IEEE Network},
  year    = {2026},
  note    = {arXiv:2511.19113},
}

@article{aip-protocol,
  title   = {{AIP}: Agent Identity Protocol for Verifiable Delegation Across {MCP} and {A2A}},
  author  = {Prakash, Shiv},
  journal = {arXiv preprint arXiv:2603.24775},
  year    = {2026},
}

@article{ldp-protocol,
  title   = {{LDP}: An Identity-Aware Protocol for Multi-Agent {LLM} Systems},
  author  = {Prakash, Shiv},
  journal = {arXiv preprint arXiv:2603.08852},
  year    = {2026},
}

@article{mcp-production,
  title   = {Bridging Protocol and Production: Design Patterns for Deploying {AI} Agents with Model Context Protocol},
  author  = {Srinivasan, Vivek},
  journal = {arXiv preprint arXiv:2603.13417},
  year    = {2026},
}

@article{iam-agentic-ai,
  title   = {Identity Management for Agentic {AI}: The New Frontier of Authorization, Authentication, and Security},
  author  = {South, Tobin and Nagabhushanaradhya, Sudhir and others},
  journal = {arXiv preprint arXiv:2510.25819},
  year    = {2025},
  note    = {OpenID Foundation whitepaper},
}

@article{oidc-a,
  title   = {{OpenID} Connect for Agents ({OIDC-A}) 1.0: A Standard Extension for {LLM}-Based Agent Identity and Authorization},
  author  = {Nagabhushanaradhya, Sudhir},
  journal = {arXiv preprint arXiv:2509.25974},
  year    = {2025},
}

@article{anp-whitepaper,
  title   = {Agent Network Protocol Technical White Paper},
  author  = {Chang, Gao and Lin, Eddie and Yuan, Chongyuan and Cai, Ran and Chen, Bin and Xie, Xinlin and others},
  journal = {arXiv preprint arXiv:2508.00007},
  year    = {2025},
}

@article{zt-agentic-ai,
  title   = {A Novel Zero-Trust Identity Framework for Agentic {AI}: Decentralized Authentication and Fine-Grained Access Control},
  author  = {Huang, Kevin and Narajala, Venkata Sai and Yeoh, Justin and Ross, Justin and others},
  journal = {arXiv preprint arXiv:2505.19301},
  year    = {2025},
}

@misc{know-your-agent,
  title        = {Know Your Agent: Governing {AI} Identity on the Agentic Web},
  author       = {Chaffer, TJ},
  year         = {2025},
  howpublished = {SSRN Working Paper},
  note         = {SSRN 5162127},
}

@article{did-vc-agents,
  title   = {{AI} Agents with Decentralized Identifiers and Verifiable Credentials},
  author  = {Garzon, Sandro Rodr{\'i}guez and Vaziry, Arian and Kuzu, Elif Merve and Gehrmann, Daniel E. and others},
  journal = {arXiv preprint arXiv:2511.02841},
  year    = {2025},
}

@article{ans,
  title   = {Agent Name Service ({ANS}): A Universal Directory for Secure {AI} Agent Discovery and Interoperability},
  author  = {Huang, Kevin and Narajala, Venkata Sai and Habler, Idan and others},
  journal = {arXiv preprint arXiv:2505.10609},
  year    = {2025},
  note    = {arXiv:2505.10609},
}

@incollection{agentic-ai-identity-security,
  title     = {Agentic {AI} Identity Security},
  author    = {Huang, Kevin and Hughes, Chris},
  booktitle = {Securing AI Agents},
  publisher = {Springer},
  year      = {2025},
  doi       = {10.1007/978-3-032-02130-4_3},
}

@article{authenticated-delegation,
  title   = {Authenticated Delegation and Authorized {AI} Agents},
  author  = {South, Tobin and Marro, Samuele and Hardjono, Thomas and Mahari, Robert and others},
  journal = {arXiv preprint arXiv:2501.09674},
  year    = {2025},
}

@inproceedings{authenticated-delegation-icml,
  title     = {Position: {AI} Agents Need Authenticated Delegation},
  author    = {South, Tobin and Marro, Samuele and Hardjono, Thomas and Mahari, Robert and others},
  booktitle = {ICML},
  year      = {2025},
  note      = {Position paper},
}

@article{decentralized-llm-identity,
  title   = {Decentralized Digital Identity Management for Large Language Model Agents},
  author  = {Aydeger, Abdullah and Zeydan, Engin and others},
  journal = {IEEE Communications Standards Magazine},
  year    = {2026},
}

@article{zt-unified,
  title   = {Fortifying the Agentic Web: A Unified Zero-Trust Architecture Against Logic-layer Threats},
  author  = {Huang, Kevin and Mehmood, Yasir and Atta, Haris and Huang, Jingbo and others},
  journal = {arXiv preprint arXiv:2508.12259},
  year    = {2025},
}

@article{zt-explainable,
  title   = {An Explainable Zero Trust Identity Framework for {LLMs}, {AI} Agents, and Agentic {AI} Systems},
  author  = {Bhushan, Bharat},
  journal = {EuroLexis Open Access Journal},
  year    = {2025},
}

@article{ssi-microservices,
  title   = {Agentic {AI} for Self-Sovereign Identity: A Decentralized Zero Trust Framework for Autonomous Microservices},
  author  = {Palavali, Dhanush Reddy},
  journal = {IJCMI},
  year    = {2025},
}

@inproceedings{spiffe-agents,
  title     = {{SPIFFE}-Based Zero-Trust Authentication for {AI} Agent Ecosystems},
  author    = {Pappu, Karthik and Bhushan, Bharat and Mittal, Aman},
  booktitle = {IEEE ICCA},
  year      = {2025},
}

@article{saavedra-delegation,
  title   = {Interoperable Architecture for Digital Identity Delegation for {AI} Agents with Blockchain Integration},
  author  = {Saavedra, Daniel R.},
  journal = {arXiv preprint arXiv:2601.14982},
  year    = {2026},
}

@article{identity-aware-governance,
  title   = {Identity-aware Governance for Autonomous {AI} Agents: A Framework for Enterprise Authorization, Delegation, and Audit},
  author  = {Ramachandran, Hari and Mishra, Gautam},
  journal = {SSRN},
  number  = {6439998},
  year    = {2026},
}

@inproceedings{auth-taxonomy,
  title     = {A Conceptual Framework for Authentication in Agentic {AI} Ecosystems: Protocol Analysis and Taxonomy},
  author    = {Bhushan, Bharat and Pappu, Karthik and Mittal, Aman},
  booktitle = {IEEE ICCA},
  year      = {2025},
}

@article{dolev-yao,
  author  = {Dolev, Danny and Yao, Andrew C.},
  title   = {On the Security of Public Key Protocols},
  journal = {IEEE Transactions on Information Theory},
  volume  = {29},
  number  = {2},
  pages   = {198--208},
  year    = {1983},
  doi     = {10.1109/TIT.1983.1056650},
}

@article{agentrfc,
  title   = {{AgentRFC}: Security Design Principles and Conformance Testing for Agent Protocols},
  author  = {Zheng, Shenghan and Zhang, Qifan},
  journal = {arXiv preprint arXiv:2603.23801},
  year    = {2026},
}

@article{anbiaee-threat-model,
  title   = {Security Threat Modeling for Emerging {AI}-Agent Protocols: A Comparative Analysis of {MCP}, {A2A}, Agora, and {ANP}},
  author  = {Anbiaee, Zeynab and Rabbani, Mahdi and Mirani, Mansur and Piya, Gunjan and others},
  journal = {arXiv preprint arXiv:2602.11327},
  year    = {2026},
}

@article{safe-ioa,
  title   = {Toward a Safe Internet of Agents},
  author  = {Wibowo, Juan A. and Polyzos, George C.},
  journal = {arXiv preprint arXiv:2512.00520},
  year    = {2025},
}

@article{hdp,
  title   = {{HDP}: A Lightweight Cryptographic Protocol for Human Delegation Provenance in Agentic {AI} Systems},
  author  = {Dalugoda, Asiri},
  journal = {arXiv preprint arXiv:2604.04522},
  year    = {2026},
  note    = {IETF Internet-Draft draft-helixar-hdp-agentic-delegation-00},
}

@article{blocka2a,
  title   = {{BlockA2A}: Towards Secure and Verifiable Agent-to-Agent Interoperability},
  author  = {Zou, Zhenhua and Liu, Zhuotao and Zhao, Lepeng and Zhan, Qiuyang},
  journal = {arXiv preprint arXiv:2508.01332},
  year    = {2025},
}

@misc{a2a-protocol,
  title        = {Agent-to-Agent ({A2A}) Protocol Specification},
  author       = {{Google}},
  year         = {2025},
  howpublished = {\url{https://a2a-protocol.org/latest/specification/}},
}

@misc{mcp-spec,
  title        = {Model Context Protocol ({MCP}) Specification},
  author       = {{Anthropic}},
  year         = {2025},
  howpublished = {\url{https://modelcontextprotocol.io/}},
}

@article{react,
  title   = {{ReAct}: Synergizing Reasoning and Acting in Language Models},
  author  = {Yao, Shunyu and Zhao, Jeffrey and Yu, Dian and Du, Nan and Shafran, Izhak and Narasimhan, Karthik and Cao, Yuan},
  journal = {arXiv preprint arXiv:2210.03629},
  year    = {2022},
}

@article{autogen,
  title   = {{AutoGen}: Enabling Next-Gen {LLM} Applications via Multi-Agent Conversation},
  author  = {Wu, Qingyun and Bansal, Gagan and Zhang, Jieyu and Wu, Yiran and others},
  journal = {arXiv preprint arXiv:2308.08155},
  year    = {2023},
}

@article{metagpt,
  title   = {{MetaGPT}: Meta Programming for a Multi-Agent Collaborative Framework},
  author  = {Hong, Sirui and Zhuge, Mingchen and Chen, Jiaqi and Zheng, Xiawu and others},
  journal = {arXiv preprint arXiv:2308.00352},
  year    = {2023},
}

@misc{oauth2,
  title        = {The {OAuth} 2.0 Authorization Framework ({RFC} 6749)},
  author       = {Hardt, Dick},
  year         = {2012},
  howpublished = {IETF RFC 6749},
}

@misc{oidc,
  title        = {{OpenID} Connect Core 1.0},
  author       = {{OpenID Foundation}},
  year         = {2014},
  howpublished = {\url{https://openid.net/specs/openid-connect-core-1_0.html}},
}

@misc{rfc8693,
  title        = {{OAuth} 2.0 Token Exchange ({RFC} 8693)},
  author       = {Jones, Michael B. and Campbell, Brian and Bradley, John and Sakimura, Nat},
  year         = {2020},
  howpublished = {IETF RFC 8693},
}

@misc{tls13,
  title        = {The Transport Layer Security ({TLS}) Protocol Version 1.3 ({RFC} 8446)},
  author       = {Rescorla, Eric},
  year         = {2018},
  howpublished = {IETF RFC 8446},
}

@misc{x509,
  title        = {Internet {X.509} Public Key Infrastructure Certificate and Certificate Revocation List Profile ({RFC} 5280)},
  author       = {Cooper, David and Santesson, Stefan and Farrell, Stephen and Boeyen, Sharon and Housley, Russell and Polk, Tim},
  year         = {2008},
  howpublished = {IETF RFC 5280},
}

@misc{gartner-agentic,
  title        = {Gartner Predicts Agentic {AI} Will Autonomously Resolve 80\% of Common Customer Service Issues by 2029},
  author       = {{Gartner}},
  year         = {2025},
  howpublished = {Gartner Press Release},
  url          = {https://www.gartner.com/en/newsroom/press-releases/2025-03-05-gartner-predicts-agentic-ai-will-autonomously-resolve-80-percent-of-common-customer-service-issues-without-human-intervention-by-2029},
}

@misc{agentmesh,
  title        = {{AgentMesh}: Production-Grade Trust Layer for Multi-Agent Systems},
  author       = {{Microsoft}},
  year         = {2026},
  howpublished = {\url{https://github.com/microsoft/agent-governance-toolkit/tree/main/agent-governance-python/agent-mesh}},
  note         = {Part of the Agent Governance Toolkit; SPIFFE/SVID-based identity, policy engine, A2A/MCP/IATP protocol bridge},
}

@misc{ap2-protocol,
  title        = {Agent Payments Protocol ({AP2}): Building a Secure and Interoperable Future for {AI}-Driven Payments},
  author       = {{Google}},
  year         = {2026},
  howpublished = {\url{https://github.com/google-agentic-commerce/AP2}},
}

@misc{ag-ui-protocol,
  title        = {{AG-UI}: The Agent-User Interaction Protocol},
  author       = {{CopilotKit}},
  year         = {2025},
  howpublished = {\url{https://github.com/ag-ui-protocol/ag-ui}},
}

@misc{acp-protocol,
  title        = {Agent Client Protocol ({ACP}): A Protocol for Connecting Any Editor to Any Agent},
  author       = {{Agent Client Protocol Project}},
  year         = {2025},
  howpublished = {\url{https://github.com/agentclientprotocol/agent-client-protocol}},
}

@misc{ibm-contextforge,
  title        = {{ContextForge}: An {AI} Gateway, Registry, and Proxy for {MCP}, {A2A}, and {REST}/{gRPC} {APIs}},
  author       = {{IBM}},
  year         = {2026},
  howpublished = {\url{https://github.com/IBM/mcp-context-forge}},
}

@misc{agntcy-identity,
  title        = {{AGNTCY} Identity: Onboarding, Creating, and Verifying Identities for Agents and {MCP} Servers},
  author       = {{AGNTCY}},
  year         = {2026},
  howpublished = {\url{https://github.com/agntcy/identity}},
}

@misc{billions-agent-identity,
  title        = {Verified Agent Identity: Decentralized Identity Management for {AI} Agents Using iden3},
  author       = {{BillionsNetwork}},
  year         = {2026},
  howpublished = {\url{https://github.com/BillionsNetwork/verified-agent-identity}},
}

@misc{agents-json,
  title        = {agents.json: An Open Specification for {API} and Agent Interaction Contracts},
  author       = {{Wild Card AI}},
  year         = {2025},
  howpublished = {\url{https://github.com/wild-card-ai/agents-json}},
}

@misc{openai-agents-sdk,
  title        = {{OpenAI} Agents {SDK}: A Lightweight Framework for Multi-Agent Workflows},
  author       = {{OpenAI}},
  year         = {2025},
  howpublished = {\url{https://github.com/openai/openai-agents-python}},
}

@misc{google-adk,
  title        = {Agent Development Kit ({ADK}): An Open-Source Python Toolkit for Building {AI} Agents},
  author       = {{Google}},
  year         = {2025},
  howpublished = {\url{https://github.com/google/adk-python}},
}

@misc{w3c-vc,
  title        = {Verifiable Credentials Data Model v2.0},
  author       = {{W3C}},
  year         = {2025},
  howpublished = {\url{https://www.w3.org/TR/vc-data-model-2.0/}},
  note         = {W3C Recommendation, 2025-05-15},
}
